\documentclass[%
 reprint,
 amsmath,amssymb,
 aps,pra,
floatfix,
]{revtex4-2}

\usepackage{graphicx}
\usepackage{subcaption}
\usepackage{dcolumn}
\usepackage{bm}

\def\beqn{\begin{eqnarray}}
\def\eeqn{\end{eqnarray}}

\def\hw{\hbar \omega}
\def\hw4{ \frac {\hbar \omega}{4}}

\def\uni{{\bf i}}
\DeclareMathOperator{\re}{e}

\def\Om{\Omega}

\def\b0{b_0}

\def\nonn{\nonumber\\}
\newcommand{\dbar}{\mathchar'26\mkern-12mu d}

\begin{document}


\title{A Pseudo-Hermitian Hybrid Model at Finite Temperature: The Role of the Exceptional Points.}

\author{Ignacio Fushimi}
\email{ifushimi@iflp.unlp.edu.ar}
\author{Marta Reboiro}%
 \email{reboiro@fisica.unlp.edu.ar}
\affiliation{IFLP, CONICET-Department of Physics. University of La Plata. \\
diag. 113 and 63, 1900 La Plata. Argentina.
}%

\date{\today}

\begin{abstract}
We study a hybrid system formed by an ensemble of colour nitrogen-vacancy centres in diamond interacting with a superconducting flux-qubit at finite temperature. The presence of impurities in the system is modelled through pseudo-hermitian Hamiltonian, by introducing an asymmetry parameter in the interaction between the superconducting flux qubit and the ensemble of colour nitrogen-vacancy centres in diamond. We construct the exact grand partition function of the system, and from it we derive the thermodynamic quantities, e.g. entropy, internal energy, and Helmholtz free energy.  In the broken symmetry phase, we observe the existence of zeros in the partition function. This zeros are  related to the existence of complex-pair-conjugate eigenvalues with real part lying among the low levels of energy. In line with the Yang–Lee framework, these zeros in the complex plane signal phase transitions, and the proposed hybrid model exhibits transitions of first-order. To account for metastable regions in parameter space, we perform a Maxwell construction and a spinodal-decomposition analysis. We determine the critical temperature at which the first zero of the partition-function appears, as a function of the asymmetry parameter and the coupling constant of the interaction between the ensemble of colour nitrogen-vacancy centres in diamond and the superconducting flux-qubit.
We also design a Carnot cycle that traverses Exceptional Points in the broken symmetry phase for temperatures above the critical value, achieving the same efficiency as the classical Carnot cycle. 
Furthermore, we implement a Stirling cycle whose efficiency surpasses its classical counterpart, particularly when operating near Exceptional Points. Finally, we outline how the model can be scaled to 
larger Hilbert-space dimensions.
\end{abstract}

\maketitle


\section{Introduction}\label{intro}

The study of the dynamics of non-Hermitian systems has grown rapidly since the work of \cite{bender1}, both theoretically \cite{ali3,bender2} and experimentally \cite{rotter1,skin}, across a range of physical platforms. Pseudo-Hermitian Hamiltonians constitute a particular class of non-Hermitian operators \cite{ali1,ali2,ali3}. It is well known that, as a function of the model parameters, the spectrum of a pseudo-Hermitian Hamiltonian may be entirely real or may contain complex-conjugate pairs of eigenvalues. In the symmetry-unbroken phase, the eigenstates of the Hamiltonian are also eigenstates of the symmetry operator and the spectrum is real; in the broken-symmetry phase, the eigenstates are no longer eigenstates of the symmetry operator and complex-conjugate eigenvalue pairs appear. The boundary between these phases consists of the so-called exceptional points (EPs).

In recent years, the thermodynamic description of non-Hermitian systems has attracted increasing interest in different physical contexts \cite{pseudo1,jarzynskipt,pt-finiteT,skinfiniteT,deffner,violation,nori,joglekar1}. The literature ranges from foundational analyses \cite{pseudo1,jarzynskipt} of the Jarzynski equality in the symmetry-unbroken phase of pseudo-Hermitian systems to contemporary experimental tests that delineate the conditions for its validity in the broken-symmetry phase \cite{joglekar1}. Among various lines of inquiry are recent results on bounds to the quantum speed limit and on uncertainty relations in non-Hermitian systems at finite temperature \cite{speed}, as well as studies of the thermodynamics of non-Hermitian Josephson junctions \cite{super}. 

Most studies address the thermodynamics of the symmetry phase, whereas the broken-symmetry phase has been explored far less extensively \cite{mraf,qrmr,mrarxiv1}. Although the partition function of pseudo-Hermitian systems is always real \cite{pseudo1}, in the broken-symmetry phase it can vanish. A recent experiment has observed zeros of the partition function in a non-Hermitian system \cite{xue} and connected them to the Yang–Lee theory of phase transitions \cite{ylee1,ylee2}. Within this framework, \cite{ylee3} relates phase transitions to properties of the ground-state entanglement entropy. In previous work \cite{mraf}, we studied first-order phase transitions in the broken-symmetry region of a bosonic mode coupled to a bath.

In the present work, we investigate the relationship between the zeros of the partition function and the existence of exceptional points (EPs) associated with eigenvalues located in the low-energy region of the spectrum. In particular, we analyse how the zeros of the partition function depend on the complex conjugate pairs of eigenvalues that emerge as the model parameters are varied from the EPs into the broken-symmetry phase. The non-Hermitian hybrid platform we consider consists of an ensemble of nitrogen–vacancy (NV) colour centres in diamond coupled to a superconducting flux qubit (SFQ). This system has attracted sustained interest since the pioneering experiment by X. Zhu and co-workers \cite{zhu}, with numerous advances reported in subsequent years \cite{zhubis,marco,nv-int-2,nv-1,nv-int-1,nv-int-new,hybrid-10,hybrid-11,nosap17,epjd}.

The work is organised as follows. In Section \ref{forma}, we present the schematic pseudo-hermitian model used to study the temperature dependence of the system composed of NVs in interaction with a SFQ. In Subsection \ref{funparticion}, we present the exact grand partition function. In Subsection \ref{apb}, we discuss the necessary conditions under which the partition function can take zero values. In Subsection \ref{thermo}, we derive the thermodynamic quantities from it, e. g. the internal energy, the Helmholtz free energy, the entropy and the specific heat. Also, in Subsection \ref{rescaling} we introduce a rescaling scheme to deal with large dimensions of the systems. In Section \ref{results}, we present and discuss the results we have obtained. We analyse the spectra and the existence of Exceptional Points. We investigate the connection between EPs and the behaviour of the thermodynamic quantities at low temperatures, particularly the zeros of the partition function. In the non-{\cal{PT}} symmetry phase, we study the condition under which the system is metastable. We set up a Carnot cycle and a Sitirling cycle across EPs.  Conclusions and outlook are presented in Section \ref{conclusions}.

\section{Formalism}\label{forma}

We shall study the behaviour of a hybrid system formed by an ensemble of Nitrogen-Vacancy, NV$^{-}$, colour centres in diamonds (NVs) interacting with a superconducting flux-qubit (SFQ) \cite{zhu,zhubis,marco,nv-int-2,nv-1,nv-int-1,nv-int-new,hybrid-10,hybrid-11,nosap17} at finite temperature. We shall adopt the pseudo-hermitian Hamiltonian analysed in \cite{epjd}. It consists of three contributions: the term corresponding to the SFQ, the Hamiltonian of the NVs and the interaction term between them. It reads

\begin{equation}
H_{hybrid}  =  H_{SFQ}+H_{S}+H_{I}.
\label{hami}
\end{equation}

The contribution of the NVs to the Hamiltonian is given by  \cite{epjd}

\begin{eqnarray}
H_{S}  & = & D ~S^2_z + E ~ (S_x^2-S_y^2), 
\label{hnvs} 
\end{eqnarray}
$\{S_x,~S_y,~S_z\}$ stands for the components of the total collective spin of the  NV ensemble. The energy scale $D=2.878$ [GHz] corresponds to the zero-field splitting between the $m_S=0$ and $m_S=\pm 1$ of the ground-state spin triplet\cite{marco,nv-int-1,nvsT}. The second term corresponds to strain-induced effect in the ensemble \cite{marco,nv-int-1}, and it can be modelled by a Lipkin-like interaction \cite{ring}.

To extend the formalism of \cite{epjd} to finite temperature, we shall model the SFQ by the microscopic pairing Hamiltonian \cite{sfqbT3,anderson,bcs}

\begin{eqnarray}
H_{SFQ} & = & \sum_i ~ \epsilon_i s_{z i} -G~\sum_{ij} ~ s_{+ i} s_{- j},
\label{hsfq}
\end{eqnarray}
$s_{+ i}$ ($s_{-i}$) is the operator that creates(destroys) a pair of particles. They can be expressed, in terms of the su(2) algebra of spins-$1/2$, $\{s_{x i},~s_{y i},~s_{z i}\}$,
as $s_{\pm i}=s_{x i} \pm \uni s_{y i}$. As it is well known, the main characteristic of the spectrum of a superconducting system is the existence of a gap between the energy of the ground state and the first excited state. This gap allows the description of the SFQ as a two-level pseudo-spin-$1/2$ system. The gap decreases when the temperature is increased in the absence of other interactions \cite{feynman,sfqbT1,sfqbT2,sfqbT3,sfqbT4,sfqbT5,sfqbT6,sfqbT7,sfqbT8}. 

In Appendix \ref{secA1}, we review the derivation of the SFQ Hamiltonian from the microscopic pairing Hamiltonian in the BCS approximation.

From the literature \cite{zhu,nv-ct1,nv-ct2,nv-ct3,linew1,linew2}, it is well known that the coherence of the NV ensemble is affected by 
impurities in diamond, e.g. impurities formed by neutral nitrogen atoms, $P_1$ centres\cite{linew1,linew2}. To model the presence of $P_1$ centres, we shall use an asymmetric coupling interaction among the NV centres and the SFQ, that is 

\begin{eqnarray}\label{eqn: hint1}
H_{I} & = &  g  ~s_{z} (\alpha S_+ +  S_-).
\label{hint}
\end{eqnarray}
The same approximation has been proposed in \cite{asymmetry,asymmetry1,asymmetry2}.

The Hamiltonian of Eq. (\ref{hami} ) is a pseudo-hermitian operator; moreover, it is {\cal{PT}}-symmetric. Thus, the spectrum is formed by either real or complex conjugate pairs of eigenvalues. In terms of the space of model parameters, two regions can be distinguished: the exact symmetry phase and the phase with broken symmetry. The exact symmetry phase is characterised by a real spectrum, while the spectrum of the phase with broken symmetry includes complex pair-conjugate eigenvalues. The border between both phases is composed of Exceptional Points (EPs). At these points, two or more eigenvalues and their corresponding eigenstates are coalescent. 

As $H_{hybrid}$ is a non-Hermitian Hamiltonian, its eigenvectors are no longer orthogonal. A bi-orthogonal base can be constructed with the eigenfunctions of $H_{hybrid}^\dagger$. The reader is kindly referred to \cite{jmp19} for further details in constructing the bi-orthogonal base and defining a proper inner product.

In general, the grand partition function at a temperature $T$ can be obtained from the density matrix $\hat \rho = \re^{-\beta (H- \mu N)}$, with $\beta= (k_B T)^{-1}$ and being $\mu$ the chemical potential, as

\begin{eqnarray}
{\cal Z}={\rm{Tr}}_\tau (\hat \rho) =\sum_n \langle \widetilde \phi_n | \hat \rho | \widetilde \phi_n \rangle_\tau=
\sum_n \langle \widetilde \phi_n | \tau \hat \rho | \widetilde \phi_n \rangle.
\label{z}
\end{eqnarray} 
In the previous equation $|\widetilde \phi_n \rangle$ are the eigenfunctions of $H$ with eigenvalue $\widetilde E_n$. The inner product is defined by introducing a metric operator $\tau$, see \cite{jmp19} for more details.

The temperature-dependent mean value of an operator $\hat O$, is given by

\begin{eqnarray}
\langle \hat O \rangle = \frac 1 {\cal Z} ~
{\rm{Tr}}_\tau (\hat \rho~ \hat O) = \frac 1 {\cal Z} ~\sum_n \langle \widetilde \phi_n | \hat \rho ~\hat O| \widetilde \phi_n \rangle_\tau.
\label{omv}
\end{eqnarray} 

Let us briefly review the derivation of the exact grand partition function for the Hamiltonian of Eq. (\ref{hami}) \cite{dapro-lipkin,dapro-pair}. 

\subsection{The exact partition function.}\label{funparticion}

For the NVs subsystem, we write the grand partition function in terms of the sum on irreducible representations of particle-hole configurations, as follows \cite{dapro-lipkin}.
Let us assume that the model space for the spins consists of two levels, each with degeneration $2 \Omega$. Thus, in the grand canonical ensemble, the number of spins, $N_S$, varies from $0$ to $4 \Om$. 
We shall give the basis of the physical space in terms of the set of  vectors

\begin{eqnarray}
\{ \mid \epsilon_1 k_1,\epsilon_2 k_2,...,\epsilon_n k_N >\},
\end{eqnarray}
$\epsilon_i$ is the index corresponding to levels, $k$ represents
substates and $i$ reads for the partition with $i$ particles and
$n$ is the particle number of the configuration. So that,
$ \epsilon_i\in \{1,2\}, k_i \in \{1,...,2 \Omega +1\},~i\in \{1,2,...,N\}$,  
  $N\in \{1,2,..., 4\Omega\}$ and  $l \in \{0,...,\infty\}$. 
The dimension of the NVs subspace is
 $2^{4 \Omega}$.

We aimed to decompose the spin subspace into invariant and
irreducible subspaces. The distribution of a given number
of particles on two degenerate levels can be represented by
numbers $\nu_1$ and $\nu_2$, i.e. $\nu_1$ is the number of
sublevels which are occupied by particles in both the lower and
the upper levels, $\nu_2$ is the number of sublevels which are
unoccupied in the lower and upper levels. The distribution of the particles on the $2
\tau$ sublevels determines the quasispin $S$ of the
state, with $2 \tau=2 \Omega-\nu_1-\nu_2$. The number
of particles in this configuration is $N=2 (\tau+\nu_1)$. We shall
call $\Gamma_{k_1,k_2,...,k_{2 (\tau+\nu_1)}}$ the subspace of
states with $\nu_1$ occupied- and $\nu_2$ unoccupied-sublevels.
The dimension is $2^{2\tau}$. There are $(2 \Omega)!/((2 \tau)!
\nu_1! \nu_2!)$ different subspaces $\Gamma_{k_1,k_2,...,k_{2
(\tau+\nu_1)}}$. Each of these subspaces can be decomposed into
irreducible ones with multiplicities $ d_S(\tau,k)$. For $N$ and $\tau$, the multiplicity of the representation $\tau-k$ is 

\begin{eqnarray} 
{\cal D}_{S}(N,\tau,k) & = &  \frac{(2 \Omega)!}{(2 \tau)!(\frac N 2 -\tau)!(2 \Omega+\frac N 2-\tau)!} d_S(\tau,k), \nonumber \\
d_S(\tau,k) & = & \frac {(2 \tau)! (2 (\tau-k)+1)}{k! (2 \tau-k+1)!}.
\end{eqnarray} 

For the SFQ subsystem, we shall sum over the irreducible representations in terms of particle-particle and hole-hole configurations \cite{dapro-pair} of the electrons that cross the Josephson junction. Let us take the basis as
\begin{eqnarray} 
\{ |j_1~m_{11},...,j_1~m_{1 l_1},j_2~m_{21},...,j_2~m_{2 l_2} >\},
\end{eqnarray}
The physical space is the direct product of the physical space of each j-level. The number of states is $2^{2 (\Omega_1+\Omega_2)}$, $2 \Omega_i$ being the degeneration of the level $i$, $j_i=2 \Omega_i +1$. As has been shown in \cite{dapro-pair}
irreducible representation $({\bf s}, s_z)$ has multiplicity

\begin{eqnarray} 
{\cal D}({\bf s}, s_z) & = & \prod_{i=1}^2 d(\Omega_i,s_i), \nonumber \\
d(\Omega_i,s_i) & = & \sum_{\tau_{qb}}~ \frac {(\Omega_i)!}{(2 \tau_{qb})! (\Omega_i-2 \tau_{qb})!}~2^{\Omega_i- 2 \tau_{qb}} g_{qb}(\tau_{qb},s_i), \nonumber \\
g_{qb}(\tau_{qb},s_i) & = & \frac {(2 \tau_{qb})! (2 s_i +1)}{(\tau_{qb}-s_i)! (\tau_{qb}+s_i+1)!},
\end{eqnarray} 
the sum on $\tau_{qb}$ runs over $\tau_{qb}=s_i,~s_i+1,~...\leq \Omega_i$, for $s_i=0,~1/2,~1,~...,\Omega_i/2$.

Thus, the exact grand partition function can be written as

\begin{widetext}
\begin{equation}
{\cal{Z}}(\beta)=
\sum_{N,\tau,k}\;\sum_{N_{qb},s_1,s_2}{\cal D}_{S}
d_{1} d_2
\sum_n \re^{-\beta [E_n(\tau-k,N_{qb},s_1,s_2)-\mu_S N_s-\mu_{qb} N_{qb}]}. \nonumber \\
\label{zfunction}
\end{equation}
\end{widetext}

It is worth working to notice that though the partition function, $\cal{Z}$, only takes real values, due to the presence of complex-pair conjugate eigenvalues, it can take the value zero in the broken symmetry phase. 

\subsection{Zeros of the partition function.}\label{apb}

We shall analyse the appearance of zeros in the partition function. 

At low temperatures the largest contribution to the partition function, $\cal Z$, comes from the energy of the ground state, $E_0$. Let us write the partition function as

\begin{eqnarray}
{\mathcal Z} = \sum_{n=0}^{N_{max}}~g(n)  {\rm e}^{- \beta E_n},
\end{eqnarray}
being $g(n)$, the degenerancy of the state with energy $E_n$. We shall split $\cal Z$ as

\begin{eqnarray}
{\cal Z}= {\cal Z}_0 +{\cal Z}', ~~~ {\cal Z}_0= g(0) \re^{- \beta E_0}.
\end{eqnarray}
If the lowest state has energy $E_0=\varepsilon \pm \uni \gamma $:

\begin{eqnarray}
{\cal Z}= 2 g(0) {\rm e}^{- \beta E_0} \cos(\beta \gamma) +{\cal Z}'.
\label{eqb1}
\end{eqnarray}
Looking at Eq.(\ref{eqb1}), the zeros of $\cal Z$ occur when  $\beta \gamma= (2 k+1) \frac \pi 2$ and ${\cal Z}'=0$, or when $\cos(\beta \gamma)$ is negative and compensate the contribution of ${\cal Z}'$. The first scenario can only occur at very low temperature. 

\subsection{Thermodynamics}\label{thermo}

From the grand partition function, $\cal Z$, we can derive the thermodynamics of the system, i.e. we can compute the entropy, $S$, the free energy, $F$, the internal energy,  
$U$,  and the specific heat, $C_V$, as usual. The corresponding expressions are given by

\beqn 
F & = & -\frac 1 \beta  \ln ( {\cal Z}) +  \mu_{p} \langle N_{p} \rangle+ \mu_{S} \langle N_{S} \rangle,\nonn
U & = & \left. -\frac{\partial \ln \cal Z} {\partial \beta}  \right\vert_{V,\mu} +  \mu_{p} \langle N_{p} \rangle+ \mu_{S} \langle N_{S} \rangle,\nonn
S & = & \left. -\frac{\partial F} {\partial T}\right\vert_{V,\mu}, \nonn
C_V & = & \left. \frac{\partial U} {\partial T} \right\vert_{V,\mu}.
\label{termo}
\eeqn
We fix  $\mu_{p}$ and $\mu_{S}$ by imposing that $\langle N_{p} \rangle$  and $\langle N_{S} \rangle$ be equal to the number of pairs and of NVs of the system, respectively.

Additionally, we can estimate the superconducting gap as a function of the temperature, $\Delta(T)$, from

\begin{align} 
\Delta & \approx  \frac{G}{2} \sqrt{\left \langle ~ \sum_j  s_{+j} s_{-j} \right \rangle}.
\label{gap}
\end{align}

In equations (\ref{termo}), the variable $V$ stands for $\alpha$ and $g$. Work can be done on the system by varying either $\alpha$ or $g$, that is
\beqn
dF = -SdT + \dbar W, ~~ \dbar W=- p_\alpha d\alpha - p_g dg.
\label{difF}
\eeqn
We shall determine the highest temperature, as function of $\alpha$ and $g$, at which the partition function has a zero, hereafter called critical temperature, $T_c$.
For temperatures below the critical temperature, we shall study
the regions of stability by using the Maxwell construction and the spinodal decomposition \cite{spinodal,mraf}. 
Once the stability regions have been established, one can set up a thermodynamic cycle and compute its efficiency. In particular, we shall establish a Carnot cycle in various regions of interest, and we shall compare its efficiency to that of a classical Carnot cycle.

\subsection{Rescaling}\label{rescaling}
The characteristic energy of the system is determined, partly, by the dimensionality. In order to study dimension-agnostic effects on the system, we need to establish a rescaling scheme for the parameters.

Starting with the NVs, a qualitative analysis of the effect of increasing the number of NVs can be understood through a Holstein-Primakoff boson mapping \cite{klein1991boson,epjd}

\beqn
S_+ & = & b_+ \sqrt{N_S - b_+b_-}, \nonn 
\quad S_- & = &S_+^\dagger,\nonn
S_z & = & b_+b_- - \frac{N_S}{2},\quad [b,b^\dagger] = 1.
\eeqn
To leading order, the spin operators behave as $S_+ = b_+ \sqrt{N_S}$ and $S_- =b \sqrt{N_S}$.
Using this as a starting point, we can see that the Lipkin coupling constant of Eq.(\ref{hnvs}), $E$, scales with $N_S$, and the interaction intensity of Eq.(\ref{hint}), $g$, scales with $\sqrt{N_S}$, hence justifying the rescaling $E\rightarrow E/N_S$ and $g\rightarrow g/\sqrt{N_S}$. 

On the side of the SFQ, the pairing gap of Eq. (\ref{gap}) has a dependence on $N_{p}$, that being the number of electron pairs. To make this dimension-agnostic, we calculated $\Delta(T=0, G)$ for various dimensions and rescaled $G$, by fitting function, $f(N_p)$, such that $\Delta(T=0, G_0) = \Delta_0 \approx D$. Additionally, we rescaled the temperature by $\Delta_0$ to complete the scheme.

All in all, this leaves us with the following scheme, given $\Omega_1 = \Omega_2=N_p$, where $N_p$ is the number of pairs considered: 
\begin{align}\label{eqn: rescaling scheme}
G&\rightarrow G_r = G_0 f(N_p), \\
\label{gapr}
T&\rightarrow T_r = \frac{T}{\Delta_0},\\
g&\rightarrow g_r = \frac{g}{\sqrt{N_S}},\\
E&\rightarrow E_r = \frac{E}{N_S},
\end{align}
where
$
f(N_p) = \frac{2.7289}{ 2 N_p + 0.73029},
$
and $G_0 =3.006$ [GHz].


\section{Results and discussion}\label{results}

In what follows, we shall fix the value of the parameters of the NV-Hamiltonian at the values $D=2.878$ [GHz], and $E=0.26$ [GHz].  We shall choose the parameters of the SFQ so that the pairing gap is comparable to the scale of energy of the NVs at zero temperature. For simplicity, we shall assume $\Omega_1= \Omega_2$. 

The systems considered hereafter consist of $N_S= 2 \Omega$, and $N_p=  \Omega_1$, which fixes the quimical potential, $\mu$, to a constant value as a function of the temperature. We choose the Hamiltonian constants, such that $\mu_S=\mu_p=0$ [GHz].



Let us begin with the study of the spectra of $H_{hybrid}$ of Eq. (\ref{hami}).

\begin{figure}
\centering
\includegraphics[width=\linewidth]{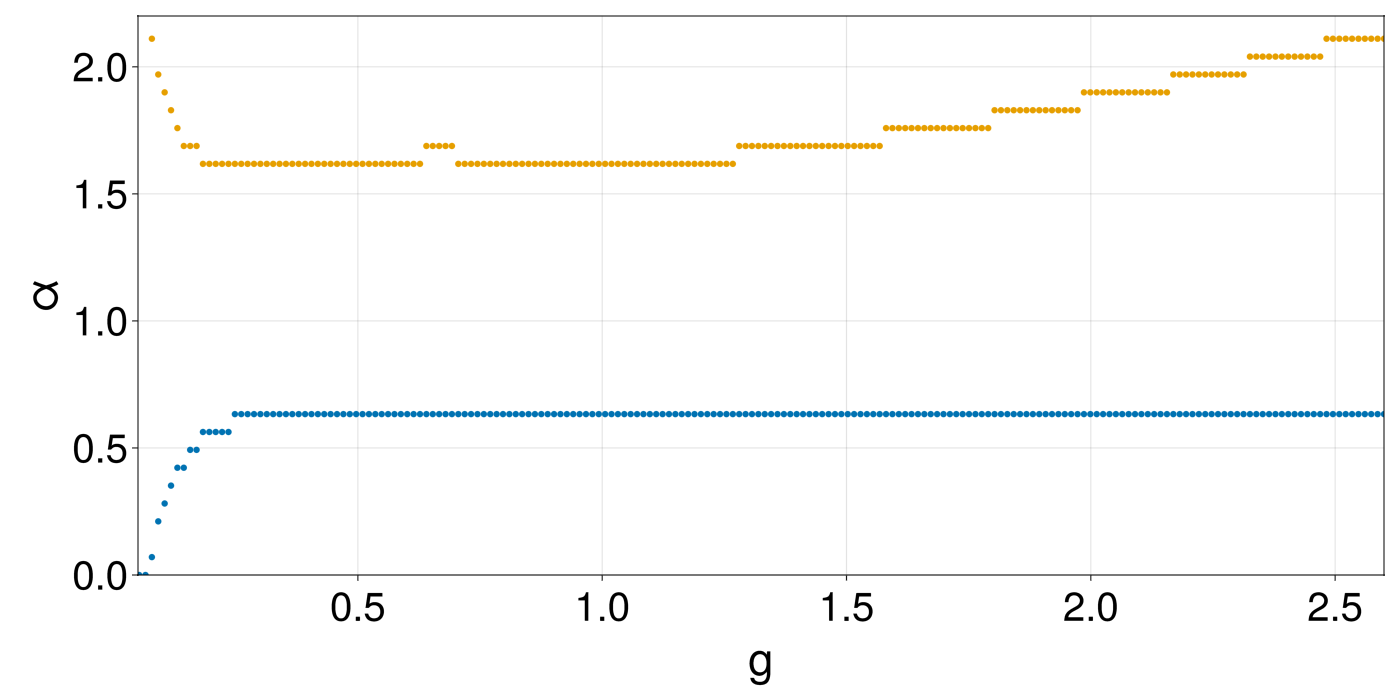}
\caption{Position of first EPs above and below $\alpha = 1$ as a function of $g$. The system consists $N_{NVs}=8$ NVs and $N_p=2$ pairs to model the SFQ. For the model parameter presented in the text, $D=2.878$ [GHz], $E=0.26$ [GHz], $G=1.73$ [GHz], $\epsilon_2=-\epsilon_1=1$ [GHz].}
\label{fig2}  
\end{figure}

The existence of EPs governs dynamical phase transitions in the system. They form the border between the region, in the model space of parameters, with real spectrum and that with complex conjugate-pair eigenvalues. The position of the EPs depends on the asymmetry parameter $\alpha$ and the coupling constant $g$. Regular patterns of EPs can be established for system with different dimensions \cite{epjd}. 
Figure \ref{fig2} depicts the position of first EPs above and below $\alpha = 1$ as a function of $g$.
We have considered a system with 
$N_{S}=8$ NVs and $N_p=2$ pairs to model the SFQ. For the model parameter presented in the text,  $G=1.73$ [GHz], $\epsilon_2=-\epsilon_1=1$ [GHz].
We can see that as the interaction coupling constant, $g$, is increased, first EPs below $\alpha=1$ tend to an almost constant value, and the values of $\alpha$ of the first EP increase almost linear with $g$ for EPs above $\alpha=1$. The value $\alpha=1$ corresponds to assuming that $H_{hybrid}$ of Eq. (\ref{hami}) is a perfect system without P1-centres. 

\begin{figure*}[t]
\begin{subfigure}[t]{0.32\textwidth}
  \includegraphics[width=\linewidth]{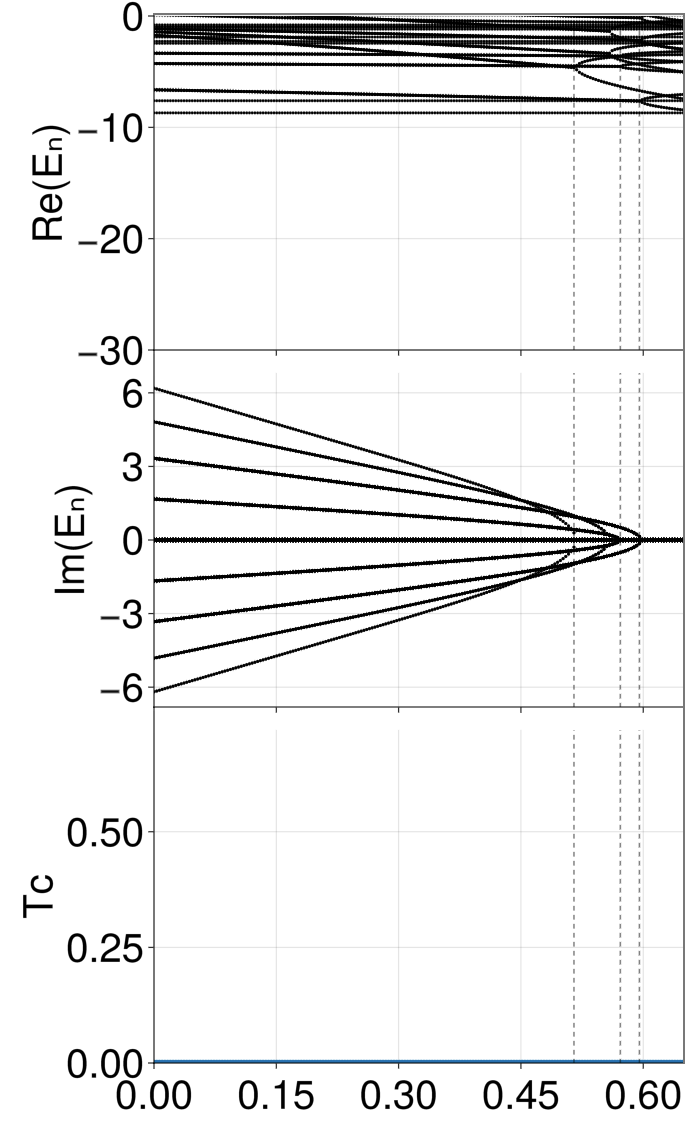}
  \caption{}
  \label{fig:1a}
\end{subfigure}\hfill
\begin{subfigure}[t]{0.32\textwidth}
  \includegraphics[width=\linewidth]{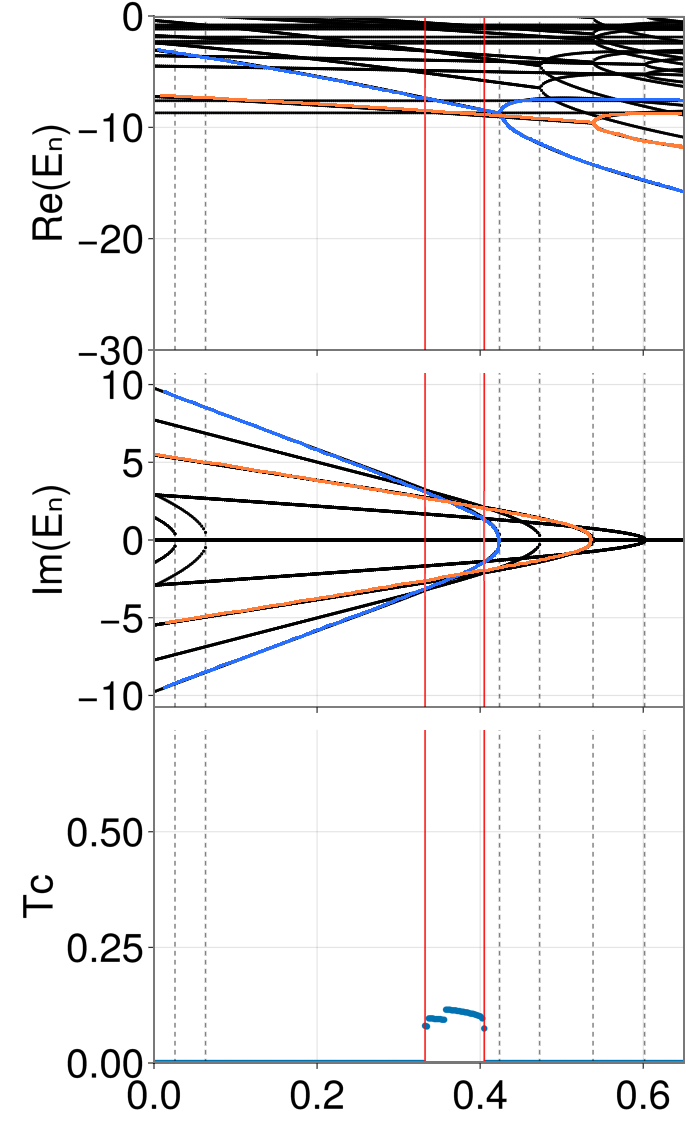}
  \caption{}
  \label{fig:1b}
\end{subfigure}\hfill
\begin{subfigure}[t]{0.32\textwidth}
  \includegraphics[width=\linewidth]{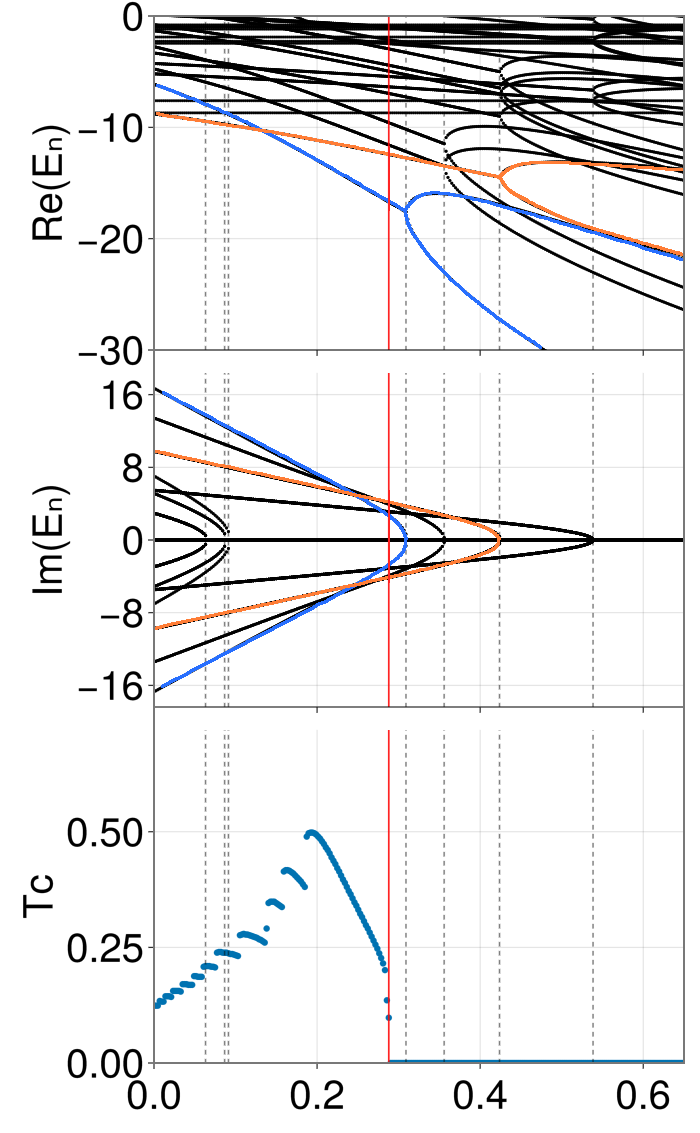}
  \caption{}
  \label{fig:1c}
\end{subfigure}
\caption{The figure shows the spectrum and the critical temperature as a function of $\alpha$ of a system of $N_S=8$ NVs and $N_p=2$ pairs to model the SFQ. For the model parameter presented in the text, $D=2.878$ [GHz], $E=0.26$ [GHz], $G=1.73$ [GHz], $\epsilon_2=-\epsilon_1=1$ [GHz].  
The upper panels display the real contribution to the eigenenergies of the system as a function of the asymmetry parameter $\alpha$. The middle panels show the imaginary part of the eigenenergies as a function of the parameter $\alpha$.  
The lower panels depict the critical temperature as a function of $\alpha$, that is, the temperature of the first zero of the partition function. Red lines are drawn to show the position of the limits of the critical domain in $\alpha$. Dotted lines correspond to the position of the EPs.  
The left column corresponds to a coupling between the NVs and the SFQ of $g=1$ [GHz], while for the middle column $g=G$, and for the right column $g=2G$, respectively.}
\label{fig1}
\end{figure*}

The thermodynamic properties of the system are closely related to the existence of EPs associated with eigenvalues located in the low-energy region of the spectrum, close to the ground state of the system. Though the grand partition function of Eq.(\ref{z}) is real\cite{jarzynskipt}, it can take zero as a possible value \cite{xue,mraf,ylee1,ylee2,ylee3}. In this context, we can introduce a critical temperature, $T_c$, it will represent the largest temperature at which the partition function takes the value zero. In Figure \ref{fig1}, we show the lower portion of the spectrum and the critical temperature, $T_c$, as a function of $\alpha$, for different values of $g$. The results correspond to 
the system described in Figure \ref{fig2}.
The upper, middle and lower panels display 
the real and imaginary parts of the eigenvalues, and criticality, respectively; all  
with respect to the asymmetry parameter $\alpha$. Red lines are drawn to show the position of the limits of the critical domain in $\alpha$. Dotted lines correspond to the position of the EPs. 
 The left column corresponds to a coupling between the NVs and the SFQ of $g=1$ [GHz], while for the middle column $g=G$, and for the right column 
$g=2~G$, respectively. There is a correspondence between the range in $\alpha$ for which the real part of a complex-pair-conjugate eigenvalue is the lowest real energy of the spectrum and the existence of zeros in the grand partition function. Also, as $g$ increases the real part of the energy of the complex eigenvalue decreases, and it becomes the ground state for  $g$ equal or larger than the pairing constant $G$. Observed that when the EPs appear to higher energy than the ground state, there are non zeros in the partition function.  

Thus, the zeros of $\mathcal{Z}$ appear at low temperatures when the ground state of the system has a complex energy. In this case, $\mathcal{Z}_0$, as defined in Eq.~(\ref{eqb1}), can take negative values, and this term compensates the contribution of $\mathcal{Z}'$, as described in Subsection~\ref{apb}. As the temperature increases, $\mathcal{Z}'$ becomes greater than $|\mathcal{Z}_0|$, and the partition function becomes strictly positive.

\begin{figure*}
 \begin{subfigure}{0.45\textwidth}
\includegraphics[width=\linewidth]{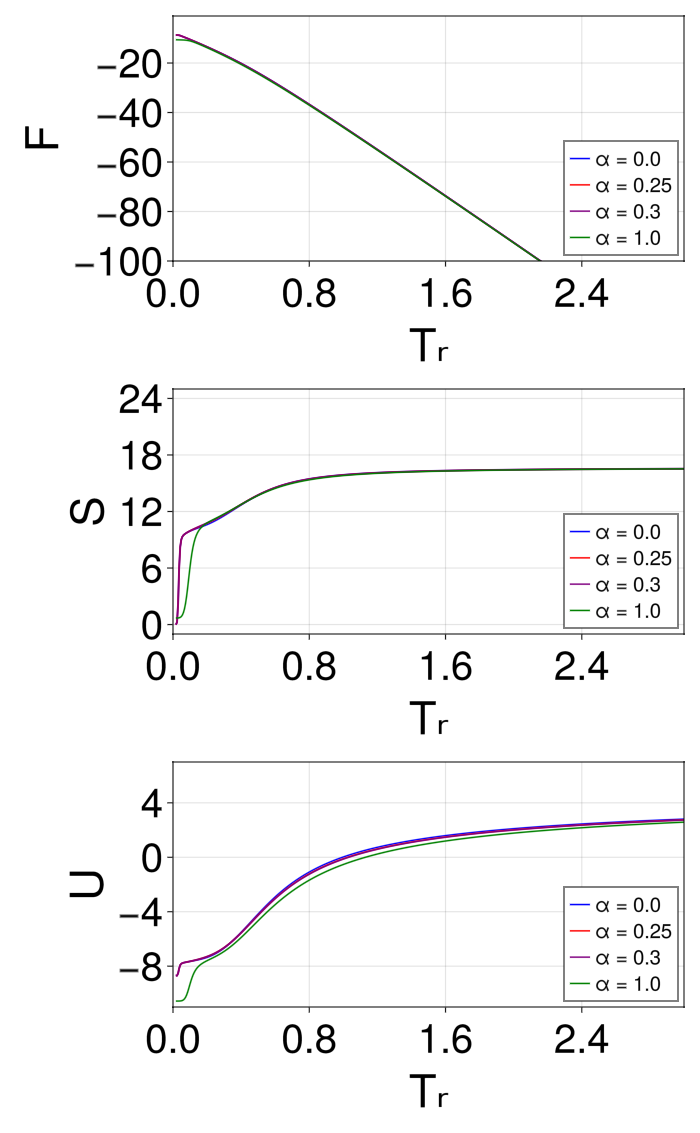}
\end{subfigure}
\begin{subfigure}{0.45\textwidth}
\includegraphics[width=\linewidth]{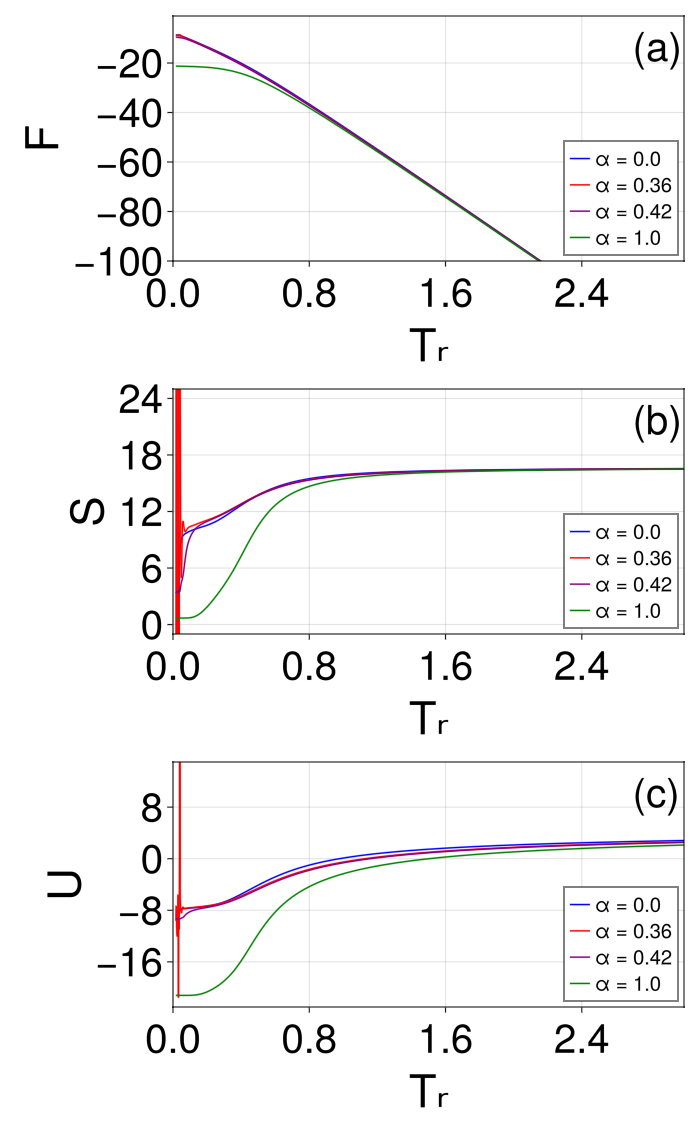}
\end{subfigure}
\caption{The figure shows the behaviour of the free energy, $F$, the entropy, $S$, and the internal energy, $U$, as a function of the temperature in units of $D$, $T_r=T/D$, in the absence of interaction between the NVs and the SFQ. The system is that of the previous Figures, $N_S=8$ NVs and $Np=2$  pairs to model the SFQ. For the model parameter presented in the text, $D=2.87$ [GHz], $E=0.26$ [GHz], $G=1.73$ [GHz], $\epsilon=1$ [GHz]. The left Panels correspond to values of the coupling constant $g=1.0$ [GHz], while the right Panels correspond to values of $g=1.73$ [GHz].}
\label{fig3}  
\end{figure*}

This fact is reflected in the behaviour of the thermodynamic quantities at temperatures below $T_c$, the aforementioned critical temperature. In Figure \ref{fig3} shows the behaviour of the free energy, $F$, the entropy, $S$, and the internal energy, $U$, as functions of the temperature in units of $D$, namely $T_r = T/D$. The system is the same as in the previous figures, with $N_S = 8$ NVs and $N_p = 2$ pairs modeling the SFQ. The left panels correspond to the coupling constant $g = 1.0$ [GHz], while the right panels correspond to $g = 1.73$ [GHz].

For the curves in the left panels, the interaction constant is fixed at $g = 1$ [GHz], i.e., $g < G$. In this case, the thermodynamic quantities are univaluated as a function of  the temperature \cite{feynman}, both the entropy and the internal energy saturate at constant values as $T$ increases, which is a signal of having a finite dimensional system. 
For the curves in the right panels, the coupling constant is comparable to $G$. In this case, for example at $\alpha = 0.36$, below the critical temperature the thermodynamic quantities display an oscillatory behaviour. 

At high temperatures, however, the asymmetry of the interaction does not affect the behaviour of the thermodynamic potentials. The asymptote value of the entropy is related to the number of configuration of the system, $S= (4 \Omega+ 2 (\Omega_1+\Omega_2)) \ln(2) $ (in units of the Boltzmann constant, $k_B=1$), and the Helmholtz free energy is linear with $T$.

As reported in \cite{mraf}, in the region below critical temperatures, the system can be interpreted as a heterogeneous one. A phase transition of first order occurs in systems for which the Helmholtz free energy has local minima separated by an energy barrier. Thus, there is a certain region, $\alpha_1 \leq \alpha \leq \alpha_2$, in which a heterogeneous state is energetically more favourable than a homogeneous one. In equilibrium conditions, the chemical potentials of the coexisting parts are equal, so that the free energy difference between both components obeys $F_1-F_2=p (\alpha_2-\alpha_1)$, with the pressure $p$ constant. 

The free energy of the heterogeneous state can be written as
\beqn
F_{het}= c_1 F_1+ c_2 F_2,
\eeqn
being $c_1$ and $c_2$ the fraction of each component in the system ($c_1+c_2=1$ and $\alpha= c_1 \alpha_1 + c_2 \alpha_2$). Using the lever rule \cite{spinodal}

\beqn
F_{het}=\frac{\alpha_2-\alpha}{\alpha_2-\alpha_1} F_1+ \frac{\alpha-\alpha_1}{\alpha_2-\alpha_1} F_2.
\label{fhet}
\eeqn
The spinodal decomposition \cite{spinodal} is done by identifying 
the minimum and the inflexion points of $F$. The extreme values of $F$ correspond to the zeros of the pressure, $p_\alpha$. The binodal zone corresponds to the interval in $\alpha$ between two minima of $F$. 
The intervals in $\alpha$ limited by a minimum and an inflexion point correspond to metastable states. In contrast, those intervals limited by inflexion points, the spinodal zone, are unstable states. The values of the Helmholtz free energy of the heterogeneous phase are obtained by connecting the minimum values of $F$ as a function of $\alpha$, Eq.(\ref{fhet}).  In this construction, the pressure $p$ is constant. It takes the absolute value of the slope of the line that connects the two successive minima of the free energy, $F$, Eq.(\ref{fhet}). Finally, the entropy can be obtained from 
$\left. \frac{\partial S}{\partial \alpha} \right \vert_{T N}=\left. \frac{\partial p}{\partial T} \right \vert_{\alpha N}$. 

\begin{figure}
\includegraphics[width=1.0\linewidth]{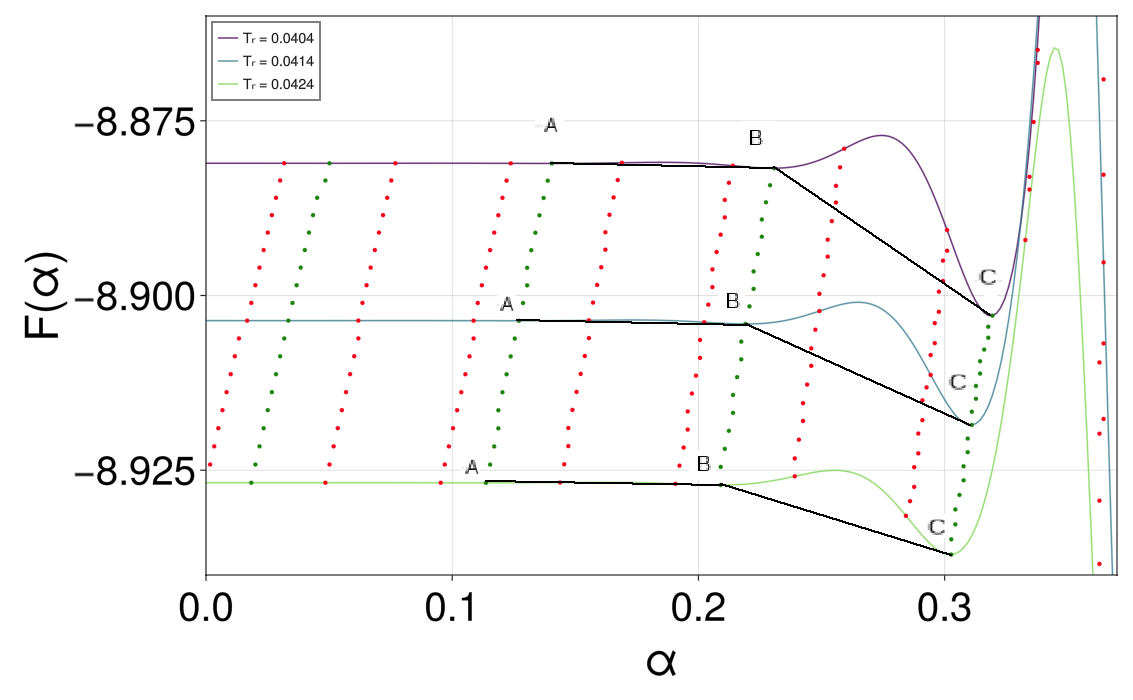}
\caption{The figure shows the essentials in the construction of the spinodal decomposition. We plot the free energy as a function of $\alpha$ for three different values of $T_r$. We draw with green dots the line that connects the minima of $F$ at different temperatures, while those connecting the inflexion points are depicted in red. The absolute value of the slopes of the black lines that connect the points A and B, and B and C, respectively, give the equilibrium pressure of the heterogeneous system.}
\label{fig4}  
\end{figure}

\begin{figure}
\includegraphics[width=\linewidth]{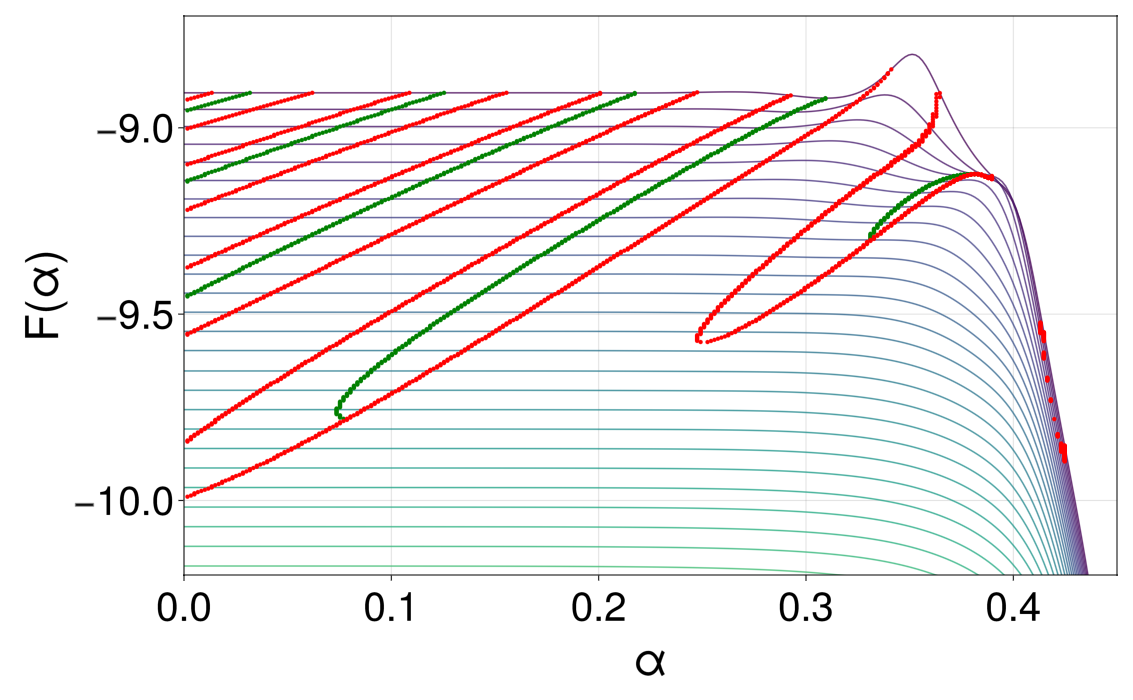}
\caption{The figure shows the behaviour of the isotherms of the Helmholtz free energy, $F$, in the range of temperatures from $T_r=0.04$ to $0.07 $, to show the evolution of minima and of inflexion points of $F$. }
\label{fig5}  
\end{figure}

In Figure \ref{fig4}, we present the spinodal construction for the system previously discussed, for coupling constant $g=1.73$ [GHz]. We plot the free energy as a function of $\alpha$ for three different values of $T_r$. We draw with green dots the line that connects the minima of $F$ at different temperatures, while those connecting the inflexion points are depicted in red. The absolute value of the slopes of the black lines that connect the points A and B, and B and C, respectively, give the equilibrium pressure of the heterogeneous system. In Figure \ref{fig5}, we have extended the number of isotherms plotted at temperatures from $T_r=0.04$ to $0.07 $, so to show the evolution of minima and of inflexion points of $F$. Above temperatures where the line of minima and the line of inflexion points coincide, the system behaves as homogeneous. 

Next, we shall discuss the possibility of the system functioning as a heat engine. We shall assume that the system is taken through equilibrium states and that the processes are reversible. 

First, we set up a Carnot cycle: the system undergoes two adiabatic processes at entropy $S_1$ and $S_2$, $S_1<S_2$, and two isothermal ones at temperatures $T_1$ and $T_2$, $T_1 < T_2$.

We have adopted the convention that both work done on the system and heat absorbed by the system are positive, while the work the system performs and the heat released are negative.

The efficiency of our system as a heat engine, $\eta$, is given in terms of the ratio of the absolute value of the net work, $W_T$, done by the system to the heat absorbed, $Q_{in}$

\beqn
\eta = \frac {|W_T|}{Q_{in}}.
\label{efi}
\eeqn
If the cycle is completed in the reverse direction, it behaves as a refrigerator, with a coefficient of performance, $e$, given by the ratio of the heat absorbed at $T_1$ to the net  work performed by the system

\beqn
e= \frac {|Q_{out}|}{W_T}.
\label{CR}
\eeqn



We will present results for the system we have studied so far, for a coupling constant of $g=1.73$ [GHz]. We have considered all possible Carnot cycles in a grid where the adopted entropies vary from $2$ to $12$, in increments of $0.5$, and for temperatures ranging from $0.1$ to $3.75$. 

We shall compare the efficiency of our cycle with the efficiency of a classical Carnot cycle

\beqn
\eta_C= 1-\frac {T_c}{T_h},
\label{efiC}
\eeqn

\begin{figure*}

  \begin{subfigure}[t]{0.48\linewidth}
    \includegraphics[width=\linewidth]{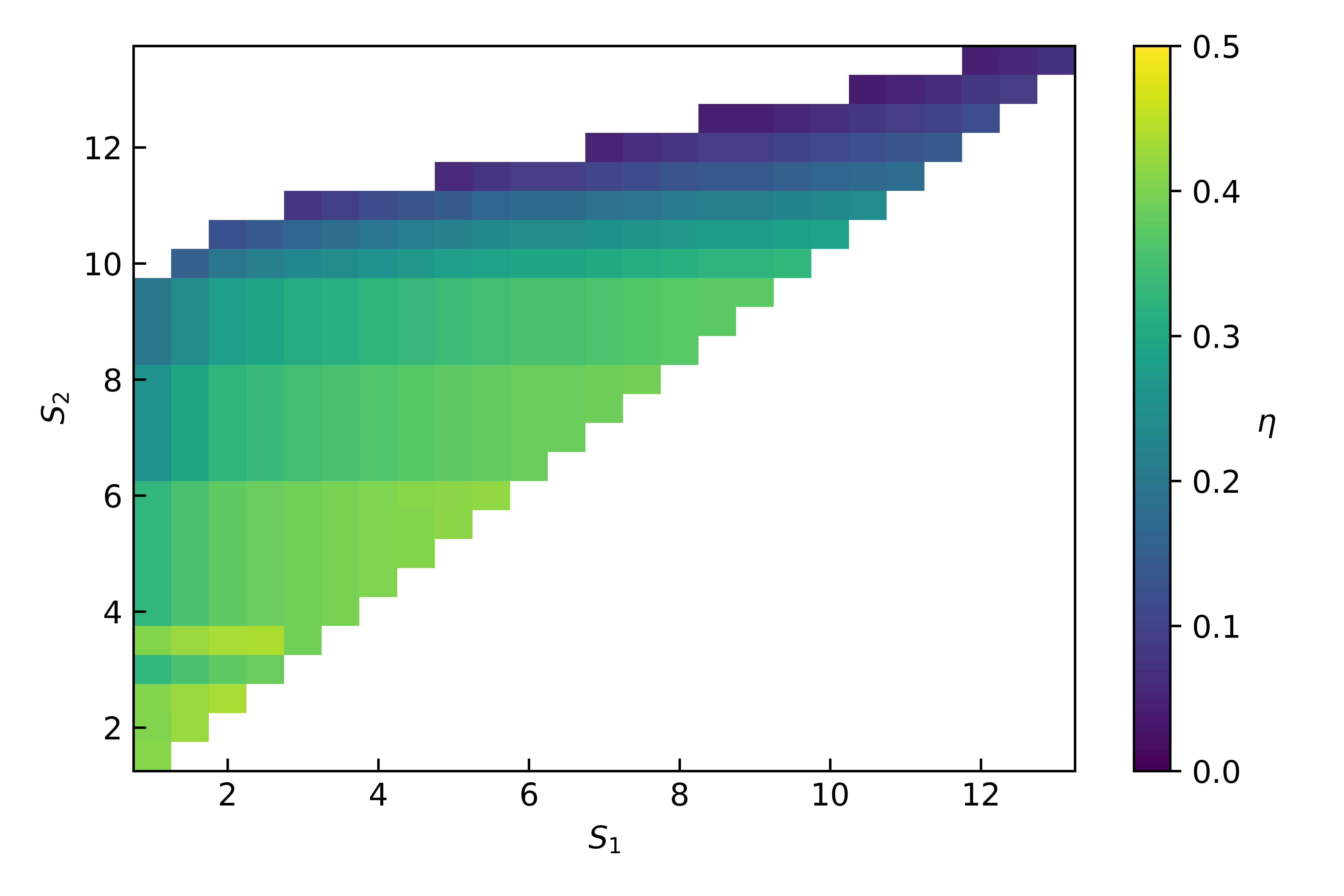}
  \end{subfigure}\hfill
  \begin{subfigure}[t]{0.48\linewidth}
    \includegraphics[width=\linewidth]{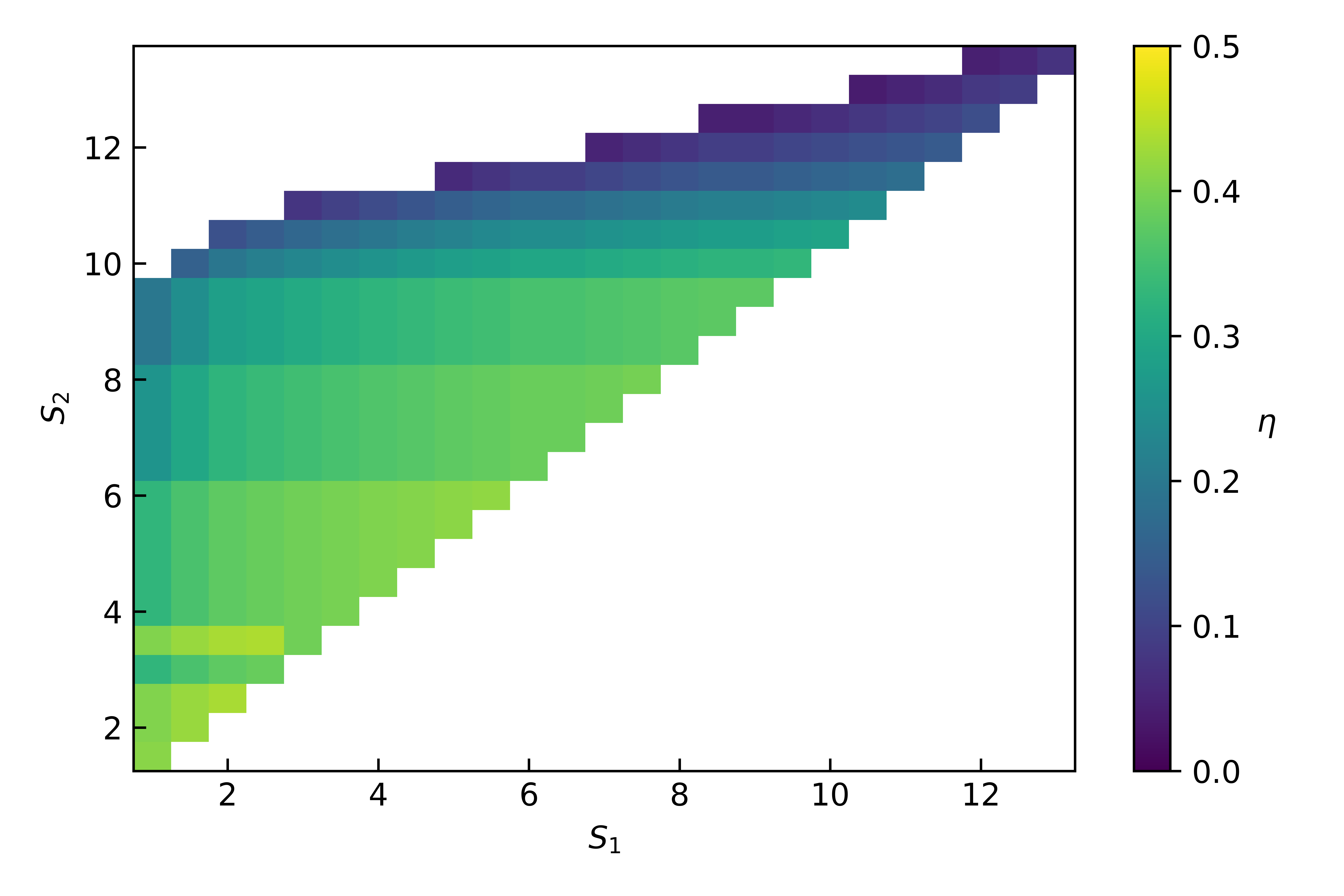}
  \end{subfigure}

  \par\medskip

  \begin{subfigure}[t]{0.48\linewidth}
    \includegraphics[width=\linewidth]{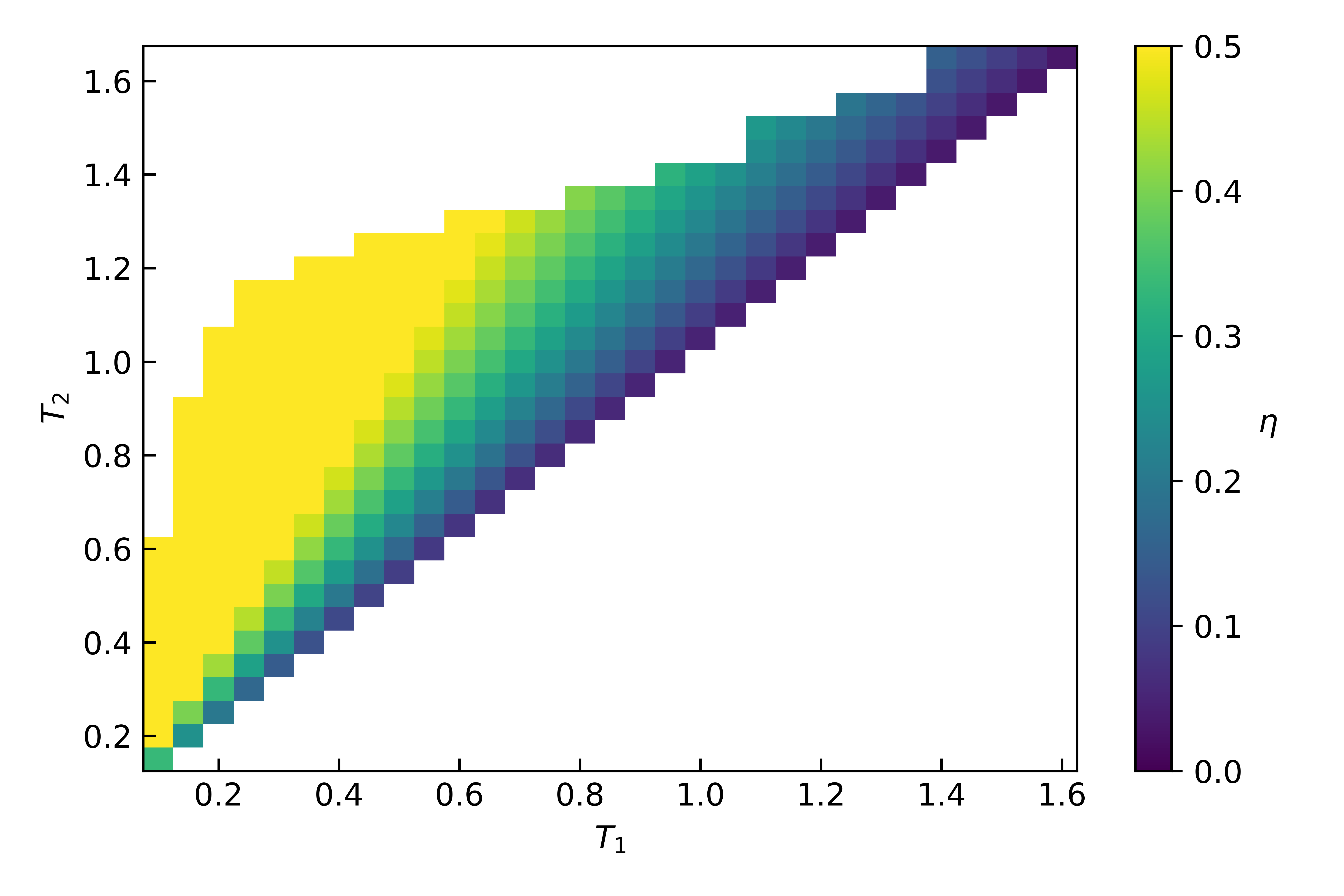}
  \end{subfigure}\hfill
  \begin{subfigure}[t]{0.48\linewidth}
    \includegraphics[width=\linewidth]{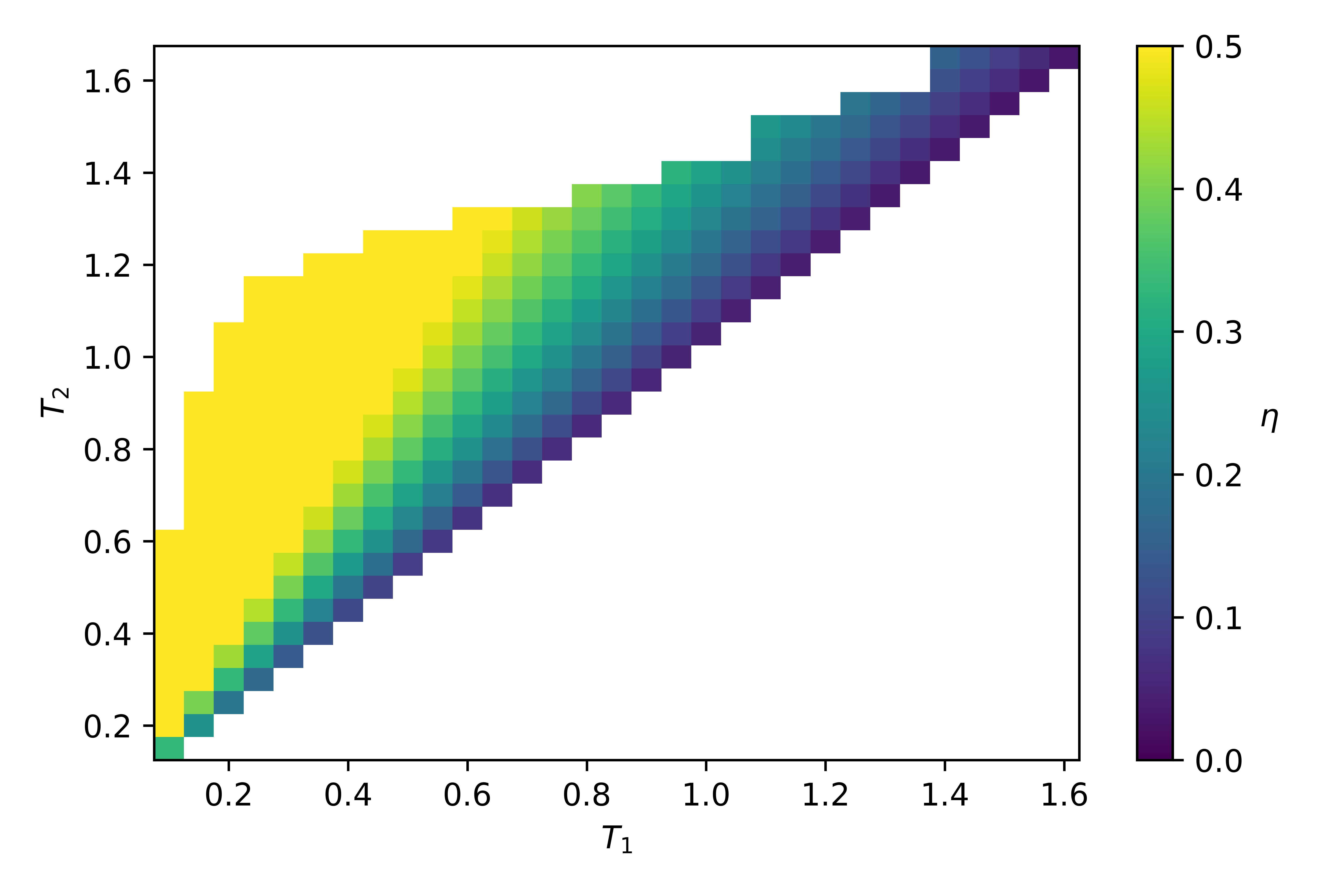}
  \end{subfigure}
   \caption{Comparison between the classical Carnot cycle and the Carnot cycle implemented in the present work. Right Column corresponds to the classical Carnot cycle, while the left column displays the results obtained for the pseudo-hermitian hybrid system. The upper row shows the efficiency that can be obtained for cycles working between entropy values $S_1$ and $S_2$. The lower row depicted the  efficiency extracted from cycles working between temperatures $T_1$ and $T_2$.  }
   \label{fig6a}
  \end{figure*}

\begin{figure*}
  \begin{subfigure}[t]{0.48\linewidth}
    \includegraphics[width=\linewidth]{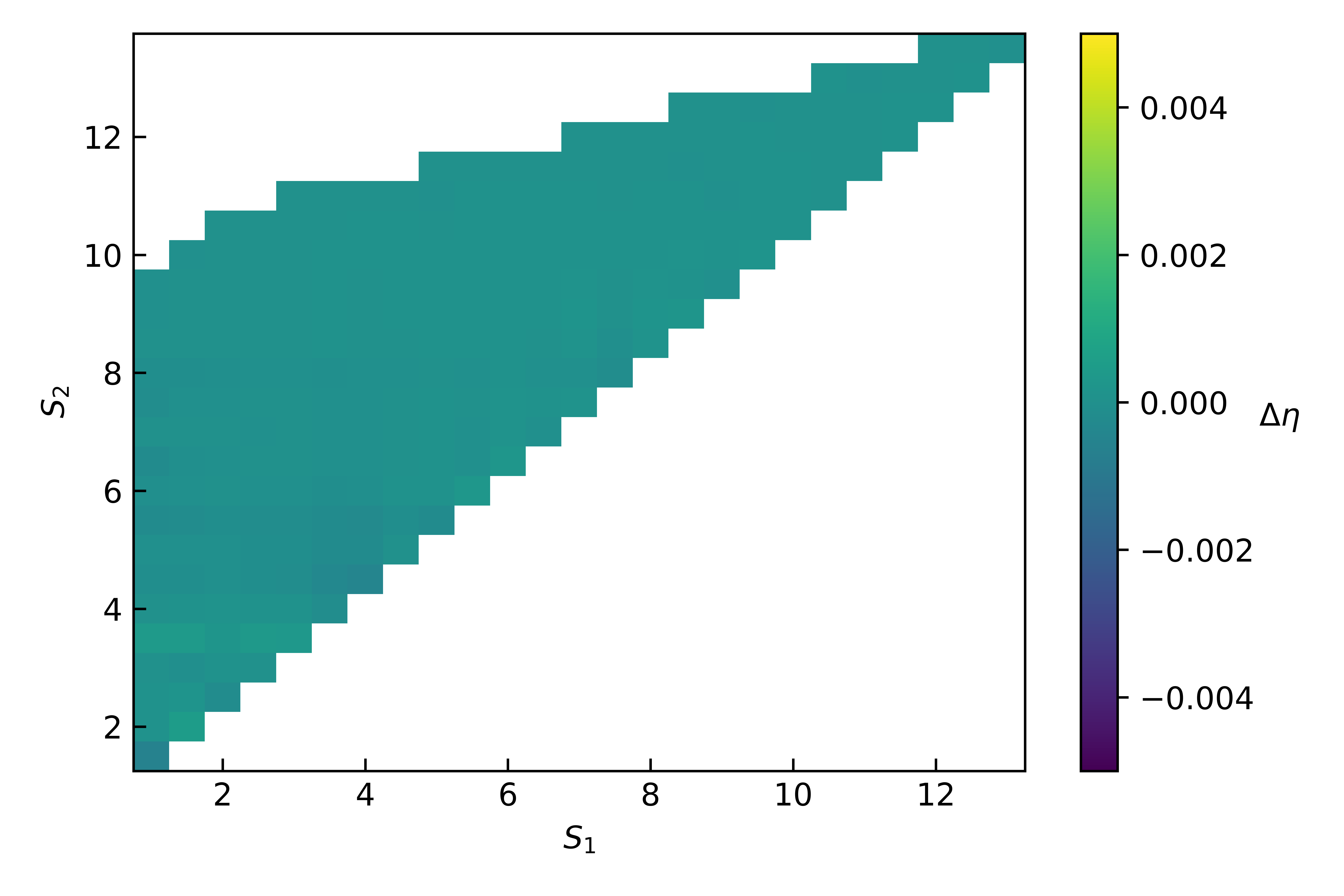}
  \end{subfigure}\hfill
  \begin{subfigure}[t]{0.48\linewidth}
    \includegraphics[width=\linewidth]{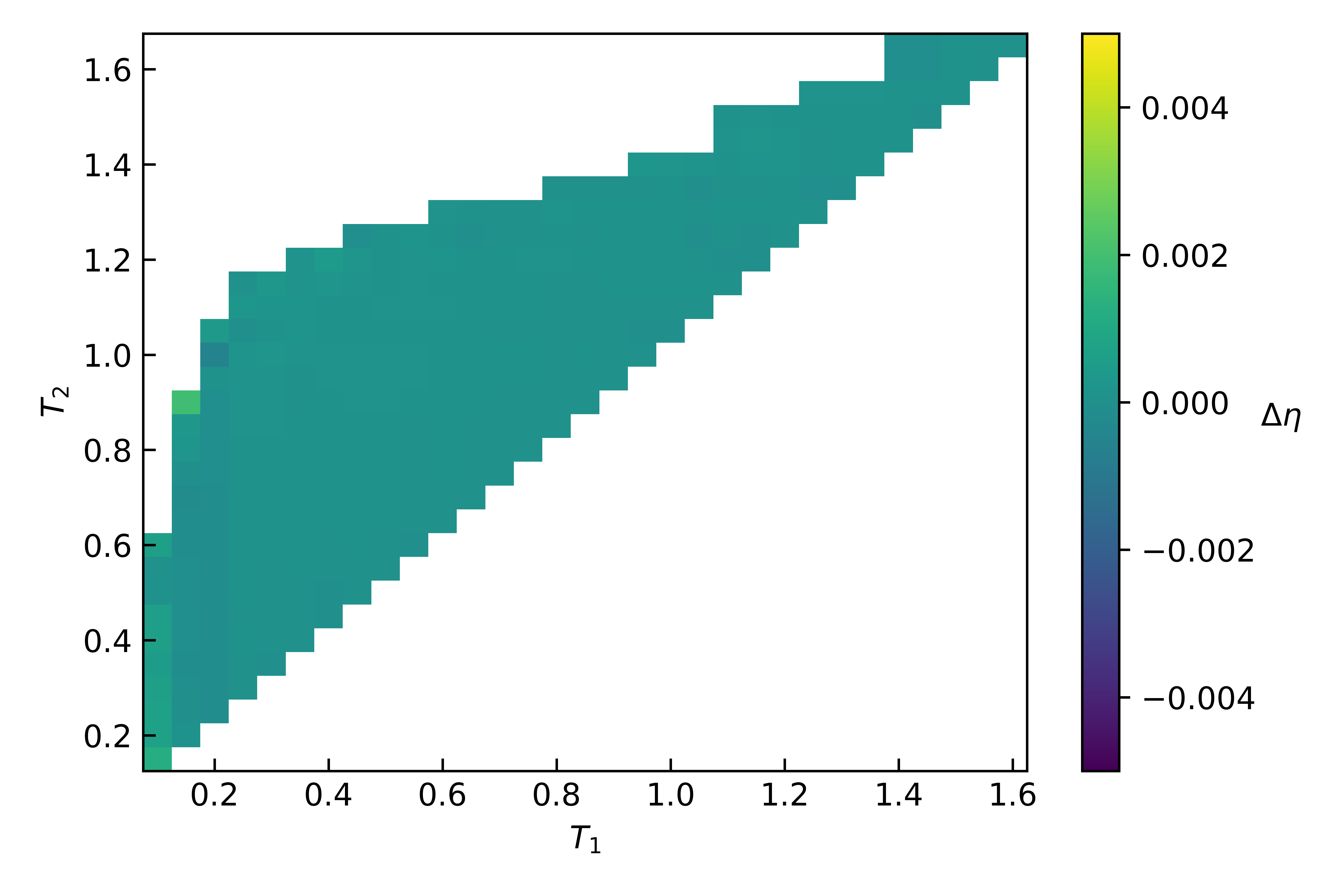}
  \end{subfigure}
  
 \caption{The difference in the efficiencies between the hybrid system cycle and classical one, both in entropy and Temperature, left and right columns respectively .  }
\label{fig6c}
\end{figure*}

In Figure \ref{fig6a}, we compare the efficiency of the classical Carnot cycle and the Carnot cycle implemented in the present work. The upper row shows the maximum efficiency that can be obtained for cycles working between entropy values $S_1$ and $S_2$. The lower row depicted the maximum efficiency extracted from cycles working between temperatures $T_1$ and $T_2$. The right column corresponds to the classical Carnot cycle, while the left column displays the results obtained for our system. As seen from the figure, the efficiency is greater for cycles working at low temperatures and intermediate entropy. In Figure \ref{fig6c}, we plot the difference in the efficiencies between the hybrid system cycle and classical one, both in entropy and temperature, left and right columns, respectively.  One of the main conclusions is that the cycle attains the classical Carnot efficiency, 
$\eta=\eta_C$, even within the broken {\cal{PT}}-symmetric phase, provided the partition function has no zeros.

\begin{figure*}

  \begin{subfigure}[t]{0.48\linewidth}
    \includegraphics[width=\linewidth]{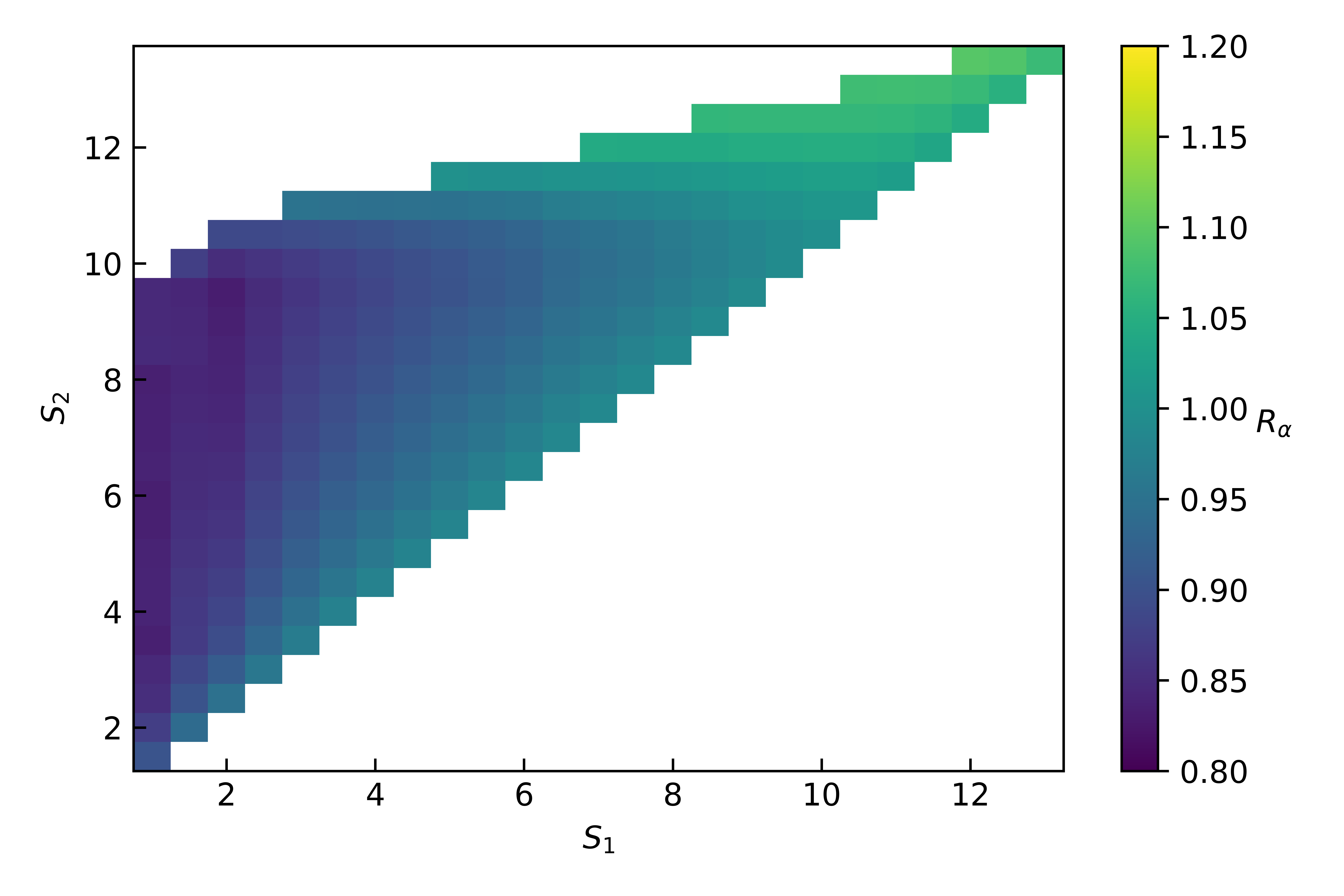}
  \end{subfigure}\hfill
  \begin{subfigure}[t]{0.48\linewidth}
    \includegraphics[width=\linewidth]{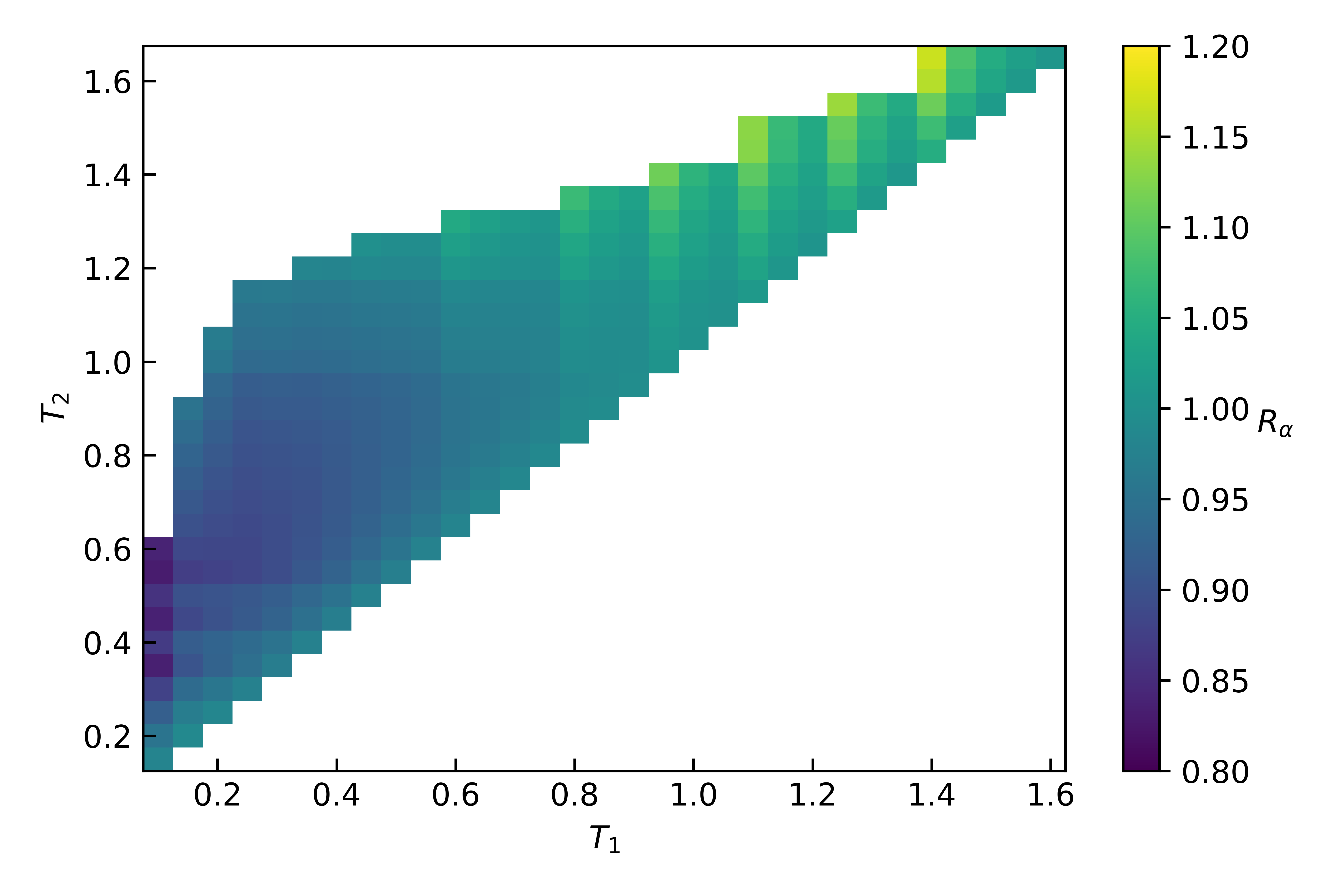}
  \end{subfigure}
  
  \caption{The figure shows the departure of the dimensionless ratio $R_\alpha=\left( \alpha_1/\alpha_2 \right)\big/\left(\alpha_4/\alpha_3\right)$ compared to the classical behaviour, $R_\alpha=1$. The value \(\alpha_i\) denotes the value of \(\alpha\) at the endpoint of leg \(i\) of the cycle.}
\label{fig6b}
\end{figure*}

In a classical Carnot cycle, the volume ratios along the two adiabatic branches are equal-i.e., \(V_1/V_2 = V_4/V_3\). Figure~\ref{fig6b} shows the departure from this behaviour by plotting the dimensionless ratio $R=\left( \alpha_1/\alpha_2 \right)\big/\left(\alpha_4/\alpha_3\right)$, where \(\alpha_i\) denotes the value of \(\alpha\) at the endpoint of leg \(i\) of the cycle. As seen in the Figure \ref{fig6b}, the departure from the classical gas is not significant.

We shall explore the construction of a Stirling cycle. That is, the system undergoes two isocoric proccesses at constant $\alpha$, $\alpha_1 < \alpha_2$, and two isothermal ones at temperatures $T_1$ and $T_2$, $T_1 < T_2$: expansion at $T_1$, absorption of heat at $\alpha_2$, compression at $T_1$, and realese of heat at $\alpha_1$. The classical efficiency of the cycle is given by

\beqn
\eta_S= \frac{T_2-T_1}
{T_2+\frac{5}{2} \frac{(T_2-T_1)} {\ln \left( \frac{\alpha_2}{\alpha_1} \right)}}.
\eeqn

\begin{figure*}
  \begin{subfigure}[t]{0.48\linewidth}
    \centering
    \includegraphics[width=\linewidth]{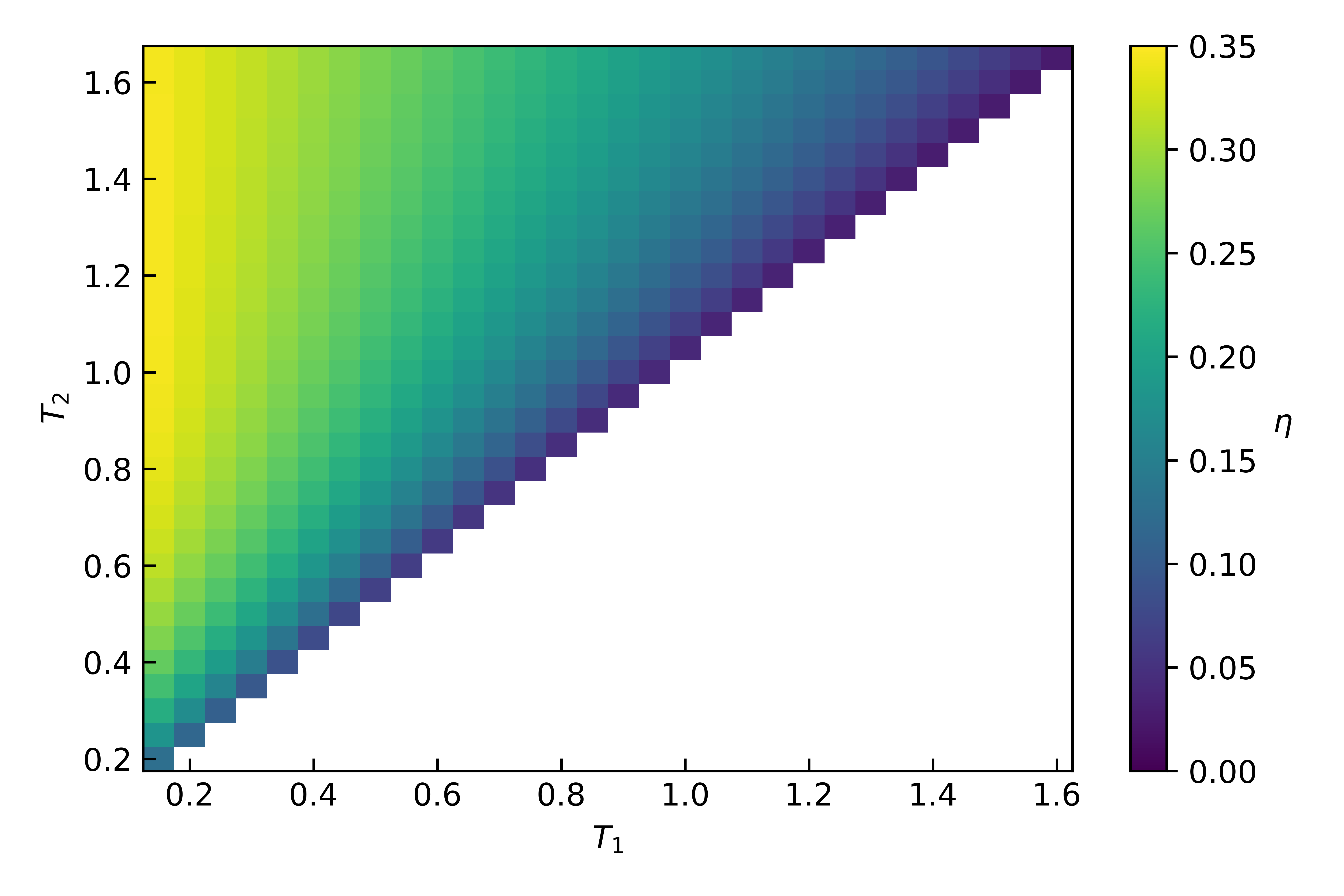}
  \end{subfigure}\hfill
  \begin{subfigure}[t]{0.48\linewidth}
    \includegraphics[width=\linewidth]{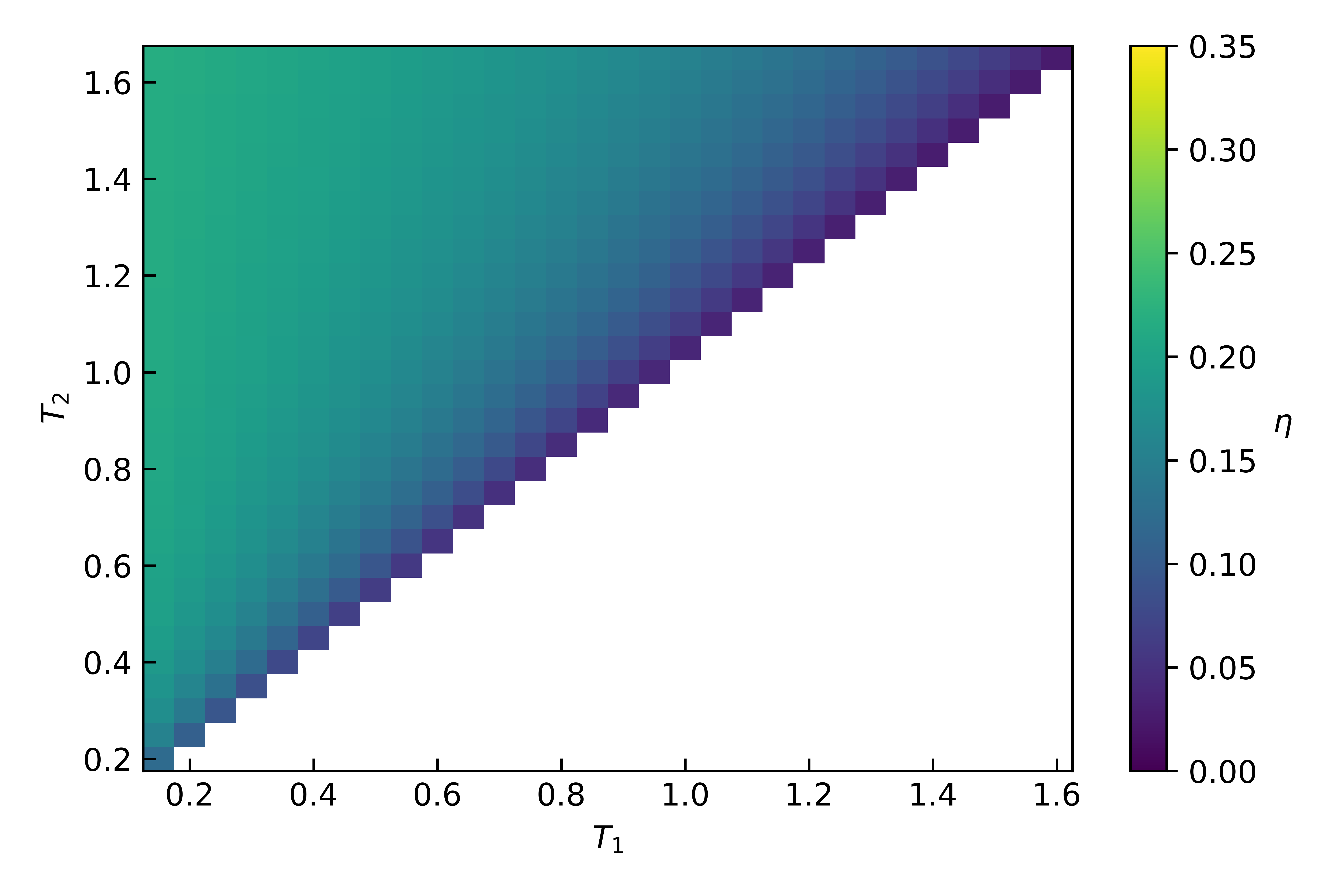}
  \end{subfigure}

  \par\medskip

  \begin{subfigure}[t]{0.48\linewidth}
    \includegraphics[width=\linewidth]{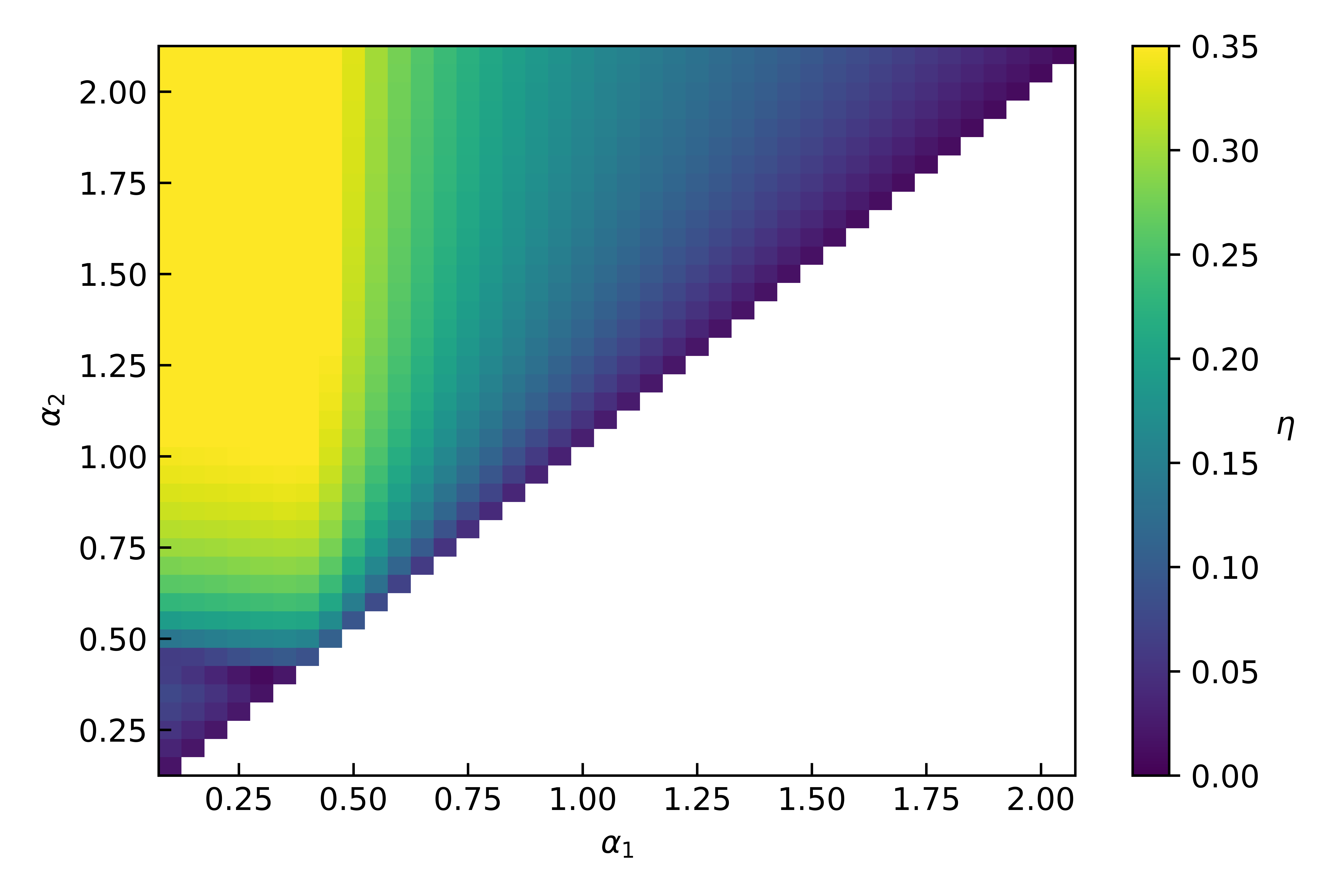}
  \end{subfigure}\hfill
  \begin{subfigure}[t]{0.48\linewidth}
    \includegraphics[width=\linewidth]{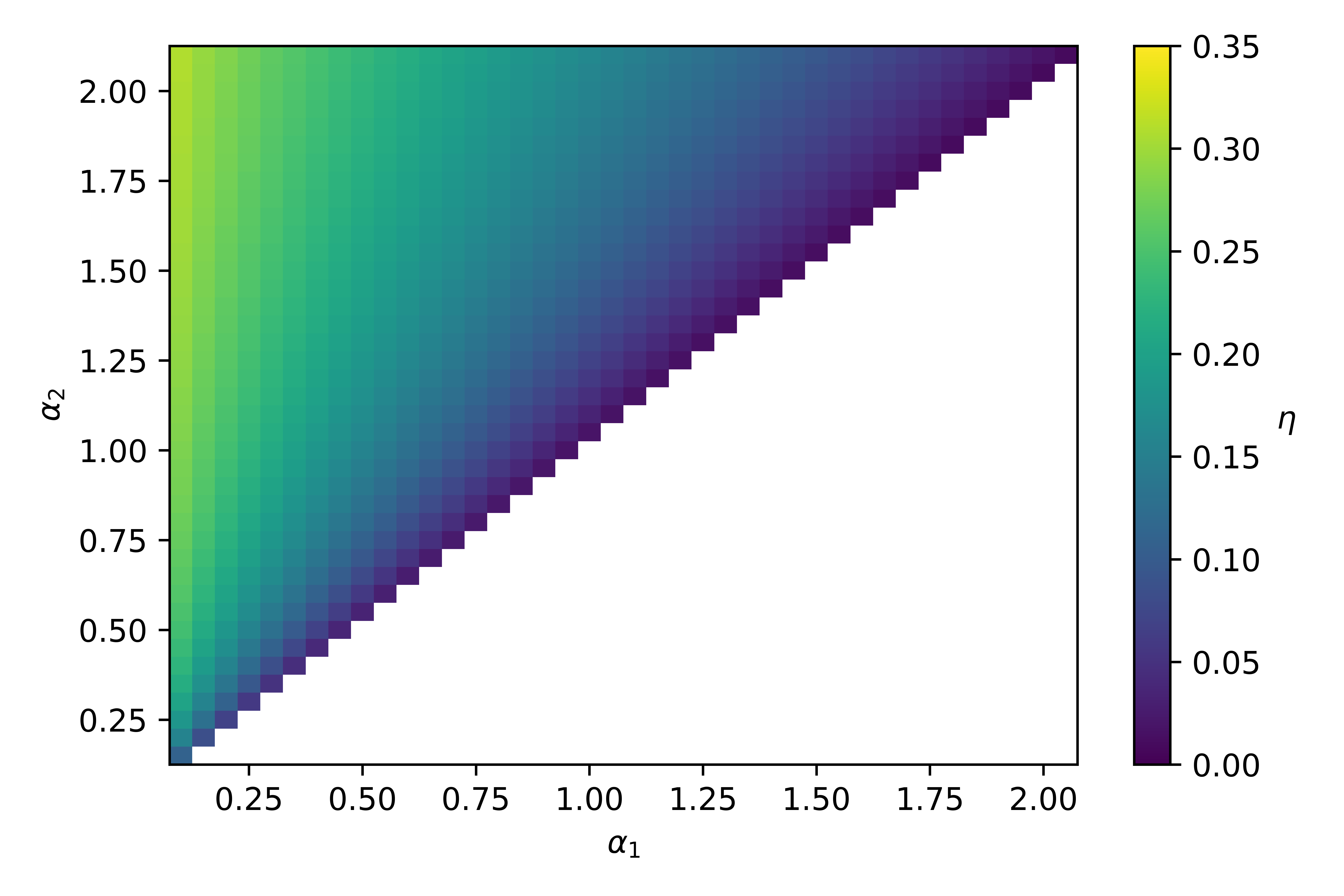}
  \end{subfigure}

\caption{Comparison between the classical Stirling cycle and the Stirling cycle implemented in the present work. the right column corresponds to the classical Stirling cycle, while the left column displays the results obtained for the pseudo-hermitian hybrid system. The upper row shows the  efficiency that can be obtained for cycles working between temperatures values $T_1$ and $T_2$. The lower row depicts the  efficiency extracted from cycles working between asymmetry constants $\alpha_1$ and $\alpha_2$.}
\label{fig7a}
\end{figure*}

\begin{figure*}
  \begin{subfigure}[t]{0.48\linewidth}
    \includegraphics[width=\linewidth]{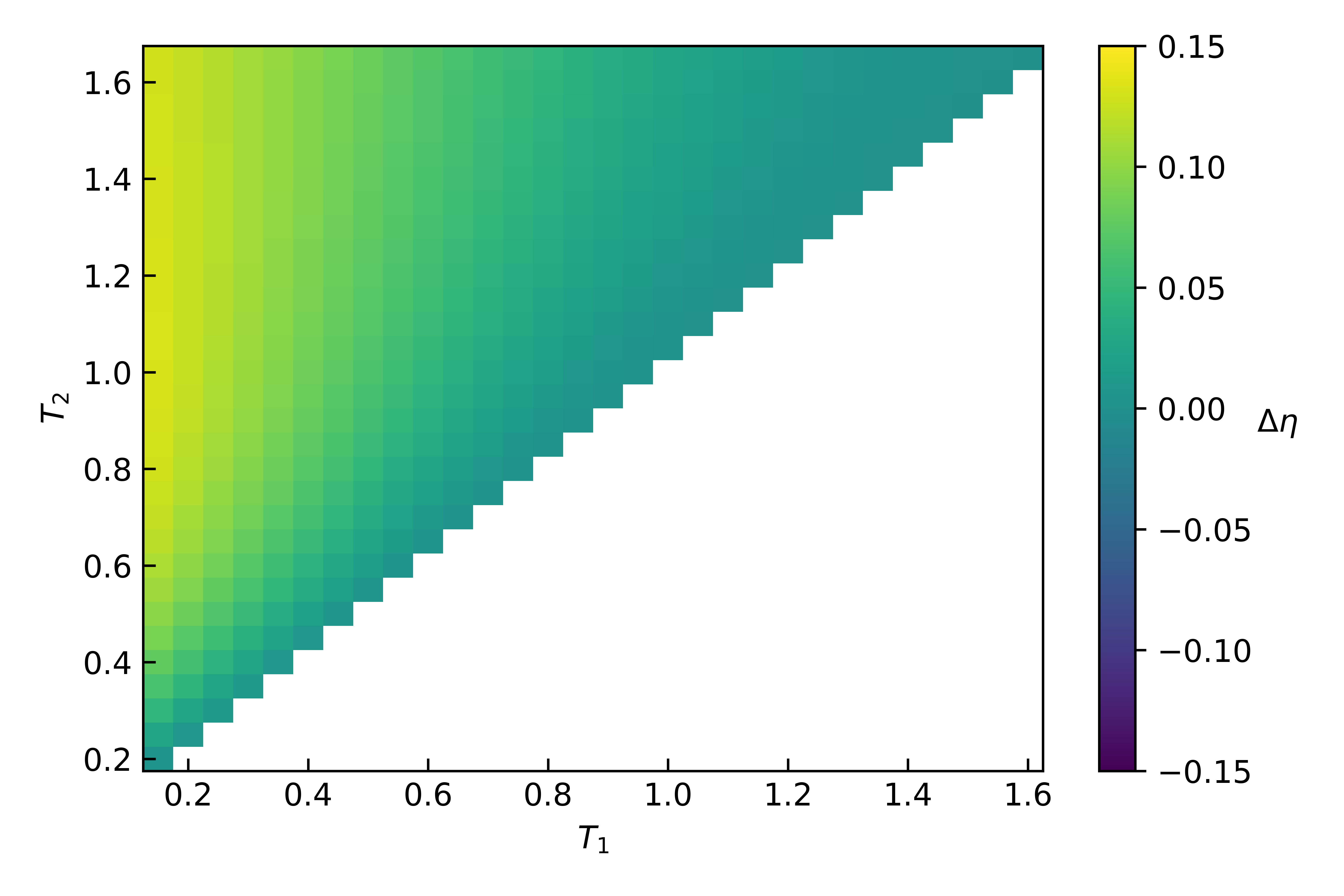}
  \end{subfigure}\hfill
  \begin{subfigure}[t]{0.48\linewidth}
    \includegraphics[width=\linewidth]{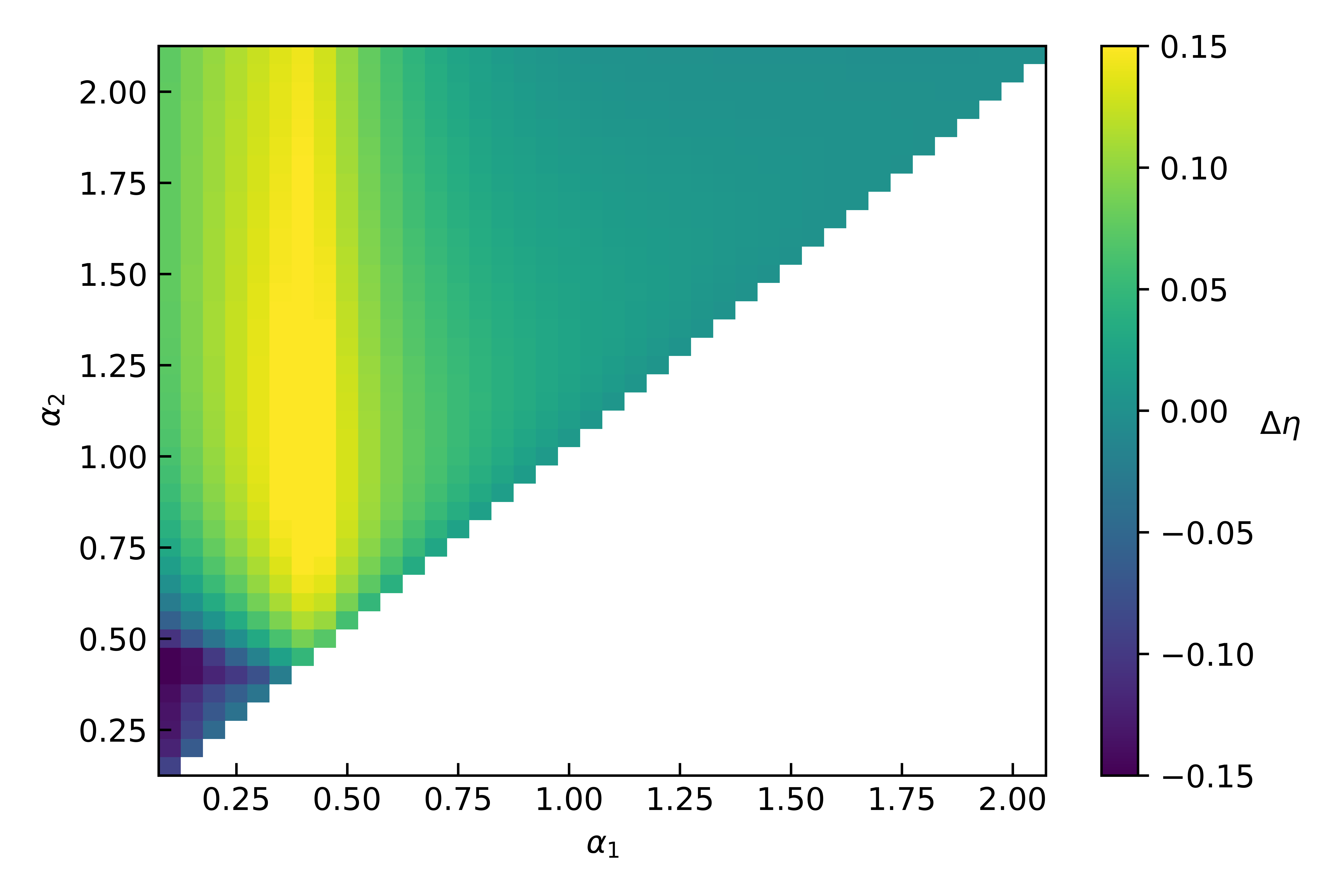}
  \end{subfigure}

\caption{The figures show the difference between the efficiency of the present cycle and the classical one, both in temperature (left Panel) and in the asymmetry constants (right Panel). }
\label{fig7b}
\end{figure*}

In Figure \ref{fig7a}, we depict the comparison between the classical Stirling cycle and the Stirling cycle implemented in the present work. The Right Column corresponds to the classical Stirling cycle, while the left column displays the results obtained for the pseudo-hermitian hybrid system. The upper row shows the efficiency that can be obtained for cycles working between temperature values $T_1$ and $T_2$. The lower row depicts the efficiency extracted from cycles working between asymmetry constants $\alpha_1$ and $\alpha_2$. Figure \ref{fig7b} shows the difference between the efficiency of the present cycle and the classical one, both in temperature and in the asymmetry constants. 

The analysis of Figures \ref{fig7a} and \ref{fig7b} indicates that, in general, the efficiency of our cycle is larger than the efficiency of the classical one. Particularly, in cycles which begin at low temperature. It should be noted that for cycles for which the lower asymmetry value, $\alpha_1$, is close to the EP with lower real energy, the cycle achieves the maximum efficiency. This can be observed if one looks at the middle Panel of Figure \ref{fig1}. 

As a final consideration, due to the collective character of the Hamiltonian of Eq. (\ref{hami}), the rescaling mechanism described in Subsection \ref{rescaling} allows us to infer the dynamics of systems with a larger number of $NVs$ and of $N_p$.

\begin{figure*}
\includegraphics[width=1.0\linewidth]{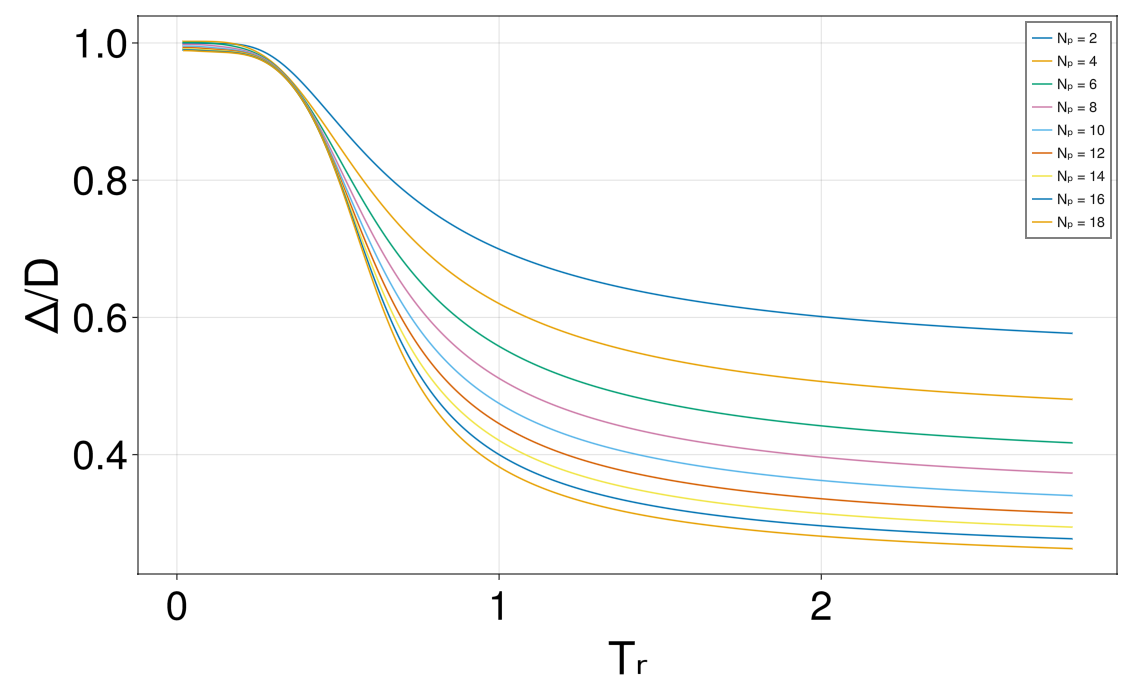}
\caption{$\Delta_r$, as function of the temperature in unit of $D$, $T_r=T/D$, for different values of $\Omega_1$, in absence of interaction with the ensemble of $NVs$. The rescale coupling constant is given in Eq. (\ref{gapr}). }
\label{fig10}  
\end{figure*}

Within the BCS formalism, the pairing gap of a superconducting system at 
$T=0$ is proportional to the number of pairs, here denoted by $\Omega_1$. 
To reduce the dependence on the system size, we derived 
Eq.~(\ref{gapr}). Figure~\ref{fig10} displays the behaviour of the rescaled 
pairing gap, $\Delta_r$, as a function of the reduced temperature, 
$T_r = T/D$, for different values of $\Omega_1$, in absence of coupling with the ensemble of $NVs$. The dimensional analysis is found to be reliable for $\Omega_1 \geq 2$ and up to temperatures of the order of $T_r=0.75$. As the number of pairs increases, the rescaling remains valid up to higher temperatures.


\begin{figure*}
\includegraphics[width=1.0\linewidth]{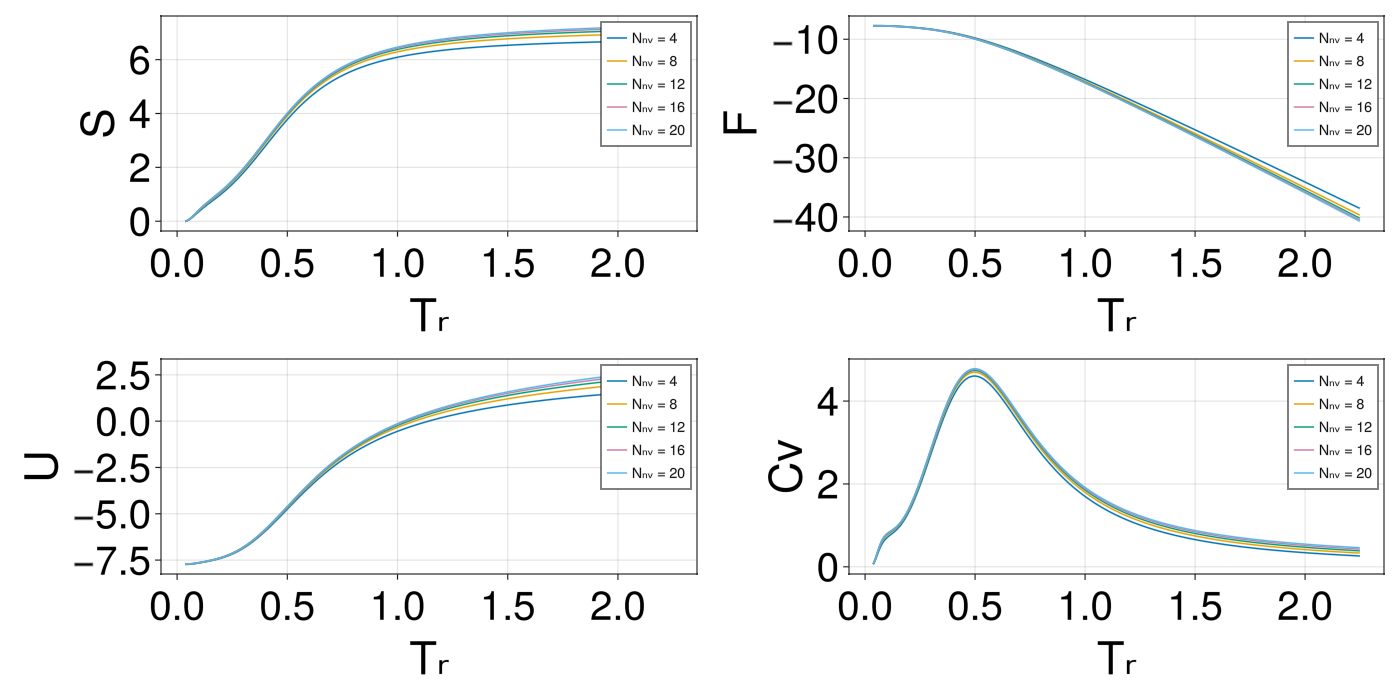}
\caption{The figure displays the behaviour of the rescaled entropy, $S$, Hemholtz free energy, $F$, internal energy, $U$ and Specific heat at constant $\alpha$, in function of the temperature in unit of $D$, $T_r=T/D$, for different mean values of the ensemble of $NVs$, $NVs=4,~8,~12,~16,~20$, in absence of interaction with the SFQ.}
\label{fig12}  
\end{figure*}

In Figure \ref{fig12}, the behaviour of the rescaled entropy, $S$, Hemholtz free energy, $F$, internal energy, $U$ and Specific heat at constant $\alpha$, in function of the temperature in unit of $D$, $T_r=T/D$, for different mean values of the ensemble of $NVs$, $NVs=4,~8,~12,~16,~20$, in absence of interaction with the SFQ. As expected, due to the collective character of the interaction, the scaling mechanism allows us to extrapolate results to systems with a large number of particles. 

\section{Conclusion}\label{conclusions}

In this work, we studied a schematic hybrid system composed of nitrogen vacancy (NVs) centres in diamond coupled to a superconducting flux qubit (SFQ), described by a pseudo-Hermitian Hamiltonian. The non-Hermitian contribution is governed by an asymmetry parameter $\alpha$ introduced to account for the presence of $P_1$ centres in the NV ensemble \cite{zhu,epjd}. We constructed the exact partition function by decomposing the model into its irreducible representations \cite{dapro-pair,dapro-lipkin,qrmr}. The spectrum is purely real in the exact {\cal{PT}}-symmetric phase, whereas complex-conjugate eigenvalues appear in the symmetry-broken phase; the boundary between these dynamical regions consists of exceptional points (EPs). The thermodynamic behaviour within the exact phase coincides with that of a Hermitian counterpart \cite{qrmr,mraf}.

To characterise the broken phase, we analysed the spectrum as a function of the asymmetry prameter $\alpha$ and the qubit–ensemble coupling $g$. For sufficiently weak $g$, the real parts of complex eigenvalues lie above the ground-state energy. Beyond a threshold in $g$, intervals of $\alpha$ emerge for which the real part of a complex eigenvalue becomes the lowest level. In that region, the ground state acquires a finite width; and the partition function, although real, can vanish. Its zeros occur precisely when the contribution of the dominant term becomes negative and cancels the contribution of the other terms. We can introduced a critical temperature, as the largest temperature at which the first zero appears for a given $\alpha$ and $g$. Within the Yang–Lee framework, such zeros signal phase transitions; accordingly, the hybrid model exhibits first-order transitions.
In the regions of temperature and asymmetry parameter, where the partition function takes zero values, the system can be interpreted as heterogeneous. 
Using a Maxwell construction and a spinodal-decomposition analysis, we obtain a Helmholtz free energy for the heterogeneous mixture that is energetically more favourable than that of the homogeneous solution. For temperatures above the critical value, we establish a Carnot cycle that traverses EPs in the non-{\cal{PT}}-symmetric phase and attains the same efficiency as the classical Carnot cycle. We also implement a Stirling cycle whose efficiency surpasses the classical one, particularly when operated in the vicinity of EPs. Finally, we outline how the model can be scaled to larger Hilbert-space dimensions beyond the minimal setting used here.

These results motivate a deeper investigation of pseudo-Hermitian systems in the symmetry-broken phase, including the role of fluctuation relations such as Jarzynski’s equality \cite{jarzynskipt,jarzynski2}.

\begin{acknowledgments}
The authors have been partially supported by grants 11/X982 from the University of La Plata (Argentina) and PIP 0457 from CONICET (Argentina). 
\end{acknowledgments}

\appendix

\section{SFQ Hamiltonian}\label{secA1}

Let us briefly review the derivation of the SFQ Hamiltonian from the microscopic pairing Hamiltonian in the BCS mean field approximation \cite{sfqbT3}.

The pairing Hamiltonian can be written as 
\begin{eqnarray}
H & = & \sum_k \epsilon_k \sigma_{z k} -G \sum_{k,k'} ~(\sigma_{x k} \sigma_{y k'}+ \sigma_{yk} \sigma_{x k'}).
\label{hp}
\end{eqnarray}
In terms of the creation (annihilation) operator of an electron of momentum $k$, $c_k^\dag (c_k)$, the Pauli operators read

\beqn
\sigma_{xk}& = & c_k^\dag c_{-k}^\dag+ c_{-k} c_k, \nonumber \\ 
\sigma_{yk}& = & -\uni (c_k^\dag c_{-k}^\dag-c_{-k} c_k), \nonumber \\
\sigma_{zk}& = &-\uni [ \sigma_{xk},\sigma_{y k}].
\eeqn
The BCS ground state \cite{bcs,anderson} 
can be written in terms of the amplitude of probability that a pair state is empty, $u_k$, or filled, $v_k$ ( with  $u_k^2 + v_k^2 = 1$), and $\phi$, which is related to the phase difference of the pair states across the Josephson Junction of the SFQ
 
\beqn
\Psi_k= \prod_k~(u_k + \re^{\uni \phi} v_k c^\dag_k c^\dag_{-k})|0 \rangle.
\eeqn

The mean-field Hamiltonian can be obtained straightforwardly by taking the mean values of $\sigma_{xk}$ and $\sigma_{yk}$ in the BCS ground state

\begin{eqnarray}
H_{MF} & = & \sum_k \epsilon_k \sigma_{z k} -G \sum_{k,k'} ~\left ( \langle \sigma_{x k} \rangle \sigma_{y k'}+ \langle \sigma_{y k} \rangle \sigma_{x k'} \right) 
\nonumber \\
& = & \sum_k \epsilon_k \sigma_{zk} + \sum_{k} ~\left ( \Delta \cos(\phi) \sigma_{xk}+ \Delta \sin(\phi) \sigma_{y k}\right),
\label{hp1}
\end{eqnarray}
where $\Delta$ is the energy gap obtained, at temperature zero, from the equation

\beqn
1=G \sum_k \frac {1}{2 E_k}, ~~~E_k=\sqrt{\Delta^2+ \epsilon_k^2}.
\eeqn
The energy of the ground state is $-E_k$, and the first excited paired state corresponds to energy $+E_k$.

%


\begin{thebibliography}{62}%
\makeatletter
\providecommand \@ifxundefined [1]{%
 \@ifx{#1\undefined}
}%
\providecommand \@ifnum [1]{%
 \ifnum #1\expandafter \@firstoftwo
 \else \expandafter \@secondoftwo
 \fi
}%
\providecommand \@ifx [1]{%
 \ifx #1\expandafter \@firstoftwo
 \else \expandafter \@secondoftwo
 \fi
}%
\providecommand \natexlab [1]{#1}%
\providecommand \enquote  [1]{``#1''}%
\providecommand \bibnamefont  [1]{#1}%
\providecommand \bibfnamefont [1]{#1}%
\providecommand \citenamefont [1]{#1}%
\providecommand \href@noop [0]{\@secondoftwo}%
\providecommand \href [0]{\begingroup \@sanitize@url \@href}%
\providecommand \@href[1]{\@@startlink{#1}\@@href}%
\providecommand \@@href[1]{\endgroup#1\@@endlink}%
\providecommand \@sanitize@url [0]{\catcode `\\12\catcode `\$12\catcode
  `\&12\catcode `\#12\catcode `\^12\catcode `\_12\catcode `\%12\relax}%
\providecommand \@@startlink[1]{}%
\providecommand \@@endlink[0]{}%
\providecommand \url  [0]{\begingroup\@sanitize@url \@url }%
\providecommand \@url [1]{\endgroup\@href {#1}{\urlprefix }}%
\providecommand \urlprefix  [0]{URL }%
\providecommand \Eprint [0]{\href }%
\providecommand \doibase [0]{https://doi.org/}%
\providecommand \selectlanguage [0]{\@gobble}%
\providecommand \bibinfo  [0]{\@secondoftwo}%
\providecommand \bibfield  [0]{\@secondoftwo}%
\providecommand \translation [1]{[#1]}%
\providecommand \BibitemOpen [0]{}%
\providecommand \bibitemStop [0]{}%
\providecommand \bibitemNoStop [0]{.\EOS\space}%
\providecommand \EOS [0]{\spacefactor3000\relax}%
\providecommand \BibitemShut  [1]{\csname bibitem#1\endcsname}%
\let\auto@bib@innerbib\@empty
\bibitem [{\citenamefont {Bender}\ and\ \citenamefont
  {Boettcher}(1998)}]{bender1}%
  \BibitemOpen
  \bibfield  {author} {\bibinfo {author} {\bibfnamefont {C.~M.}\ \bibnamefont
  {Bender}}\ and\ \bibinfo {author} {\bibfnamefont {S.}~\bibnamefont
  {Boettcher}},\ }\href {https://doi.org/10.1103/PhysRevLett.80.5243}
  {\bibfield  {journal} {\bibinfo  {journal} {Phys. Rev. Lett.}\ }\textbf
  {\bibinfo {volume} {80}},\ \bibinfo {pages} {5243} (\bibinfo {year}
  {1998})}\BibitemShut {NoStop}%
\bibitem [{\citenamefont {Ince}\ \emph {et~al.}(2025)\citenamefont {Ince},
  \citenamefont {Mermer},\ and\ \citenamefont {Mostafazadeh}}]{ali3}%
  \BibitemOpen
  \bibfield  {author} {\bibinfo {author} {\bibfnamefont {N.}~\bibnamefont
  {Ince}}, \bibinfo {author} {\bibfnamefont {H.}~\bibnamefont {Mermer}},\ and\
  \bibinfo {author} {\bibfnamefont {A.}~\bibnamefont {Mostafazadeh}},\ }\href
  {https://doi.org/10.1063/5.0264120} {\bibfield  {journal} {\bibinfo
  {journal} {Journal of Mathematical Physics}\ }\textbf {\bibinfo {volume}
  {66}},\ \bibinfo {pages} {092101} (\bibinfo {year} {2025})}\BibitemShut
  {NoStop}%
\bibitem [{\citenamefont {Bender}(2007)}]{bender2}%
  \BibitemOpen
  \bibfield  {author} {\bibinfo {author} {\bibfnamefont {C.~M.}\ \bibnamefont
  {Bender}},\ }\href {https://doi.org/10.1088/0034-4885/70/6/R03} {\bibfield
  {journal} {\bibinfo  {journal} {Reports on Progress in Physics}\ }\textbf
  {\bibinfo {volume} {70}},\ \bibinfo {pages} {947} (\bibinfo {year}
  {2007})}\BibitemShut {NoStop}%
\bibitem [{\citenamefont {El-Ganainy}\ \emph {et~al.}(2018)\citenamefont
  {El-Ganainy}, \citenamefont {Makris}, \citenamefont {Khajavikhan},
  \citenamefont {Musslimani}, \citenamefont {Rotter},\ and\ \citenamefont
  {Christodoulides}}]{rotter1}%
  \BibitemOpen
  \bibfield  {author} {\bibinfo {author} {\bibfnamefont {R.}~\bibnamefont
  {El-Ganainy}}, \bibinfo {author} {\bibfnamefont {K.~G.}\ \bibnamefont
  {Makris}}, \bibinfo {author} {\bibfnamefont {M.}~\bibnamefont {Khajavikhan}},
  \bibinfo {author} {\bibfnamefont {Z.~H.}\ \bibnamefont {Musslimani}},
  \bibinfo {author} {\bibfnamefont {S.}~\bibnamefont {Rotter}},\ and\ \bibinfo
  {author} {\bibfnamefont {D.~N.}\ \bibnamefont {Christodoulides}},\ }\href
  {https://doi.org/10.1038/nphys4323} {\bibfield  {journal} {\bibinfo
  {journal} {Nature Physics}\ }\textbf {\bibinfo {volume} {14}},\ \bibinfo
  {pages} {11} (\bibinfo {year} {2018})}\BibitemShut {NoStop}%
\bibitem [{\citenamefont {Liu}\ \emph {et~al.}(2023)\citenamefont {Liu},
  \citenamefont {Li}, \citenamefont {Yang}, \citenamefont {Shen}, \citenamefont
  {Yang}, \citenamefont {Hang},\ and\ \citenamefont {Ezawa}}]{skin}%
  \BibitemOpen
  \bibfield  {author} {\bibinfo {author} {\bibfnamefont {B.}~\bibnamefont
  {Liu}}, \bibinfo {author} {\bibfnamefont {Y.}~\bibnamefont {Li}}, \bibinfo
  {author} {\bibfnamefont {B.}~\bibnamefont {Yang}}, \bibinfo {author}
  {\bibfnamefont {X.}~\bibnamefont {Shen}}, \bibinfo {author} {\bibfnamefont
  {Y.}~\bibnamefont {Yang}}, \bibinfo {author} {\bibfnamefont {Z.~H.}\
  \bibnamefont {Hang}},\ and\ \bibinfo {author} {\bibfnamefont
  {M.}~\bibnamefont {Ezawa}},\ }\href
  {https://doi.org/10.1103/PhysRevResearch.5.043034} {\bibfield  {journal}
  {\bibinfo  {journal} {Phys. Rev. Res.}\ }\textbf {\bibinfo {volume} {5}},\
  \bibinfo {pages} {043034} (\bibinfo {year} {2023})}\BibitemShut {NoStop}%
\bibitem [{\citenamefont {Mostafazadeh}(2002)}]{ali1}%
  \BibitemOpen
  \bibfield  {author} {\bibinfo {author} {\bibfnamefont {A.}~\bibnamefont
  {Mostafazadeh}},\ }\href {https://doi.org/10.1063/1.1418246} {\bibfield
  {journal} {\bibinfo  {journal} {Journal of Mathematical Physics}\ }\textbf
  {\bibinfo {volume} {43}},\ \bibinfo {pages} {205} (\bibinfo {year}
  {2002})}\BibitemShut {NoStop}%
\bibitem [{\citenamefont {Mostafazadeh}(2003)}]{ali2}%
  \BibitemOpen
  \bibfield  {author} {\bibinfo {author} {\bibfnamefont {A.}~\bibnamefont
  {Mostafazadeh}},\ }\href {https://doi.org/10.1063/1.1539304} {\bibfield
  {journal} {\bibinfo  {journal} {Journal of Mathematical Physics}\ }\textbf
  {\bibinfo {volume} {44}},\ \bibinfo {pages} {974} (\bibinfo {year}
  {2003})}\BibitemShut {NoStop}%
\bibitem [{\citenamefont {Gardas~B.}(2016)}]{pseudo1}%
  \BibitemOpen
  \bibfield  {author} {\bibinfo {author} {\bibfnamefont {S.~A.}\ \bibnamefont
  {Gardas~B.}, \bibfnamefont {Deffner~S.}},\ }\href
  {https://doi.org/10.1038/srep23408} {\bibfield  {journal} {\bibinfo
  {journal} {Scientific Reports}\ }\textbf {\bibinfo {volume} {6}},\ \bibinfo
  {pages} {23408} (\bibinfo {year} {2016})}\BibitemShut {NoStop}%
\bibitem [{\citenamefont {Deffner}\ and\ \citenamefont
  {Saxena}(2015)}]{jarzynskipt}%
  \BibitemOpen
  \bibfield  {author} {\bibinfo {author} {\bibfnamefont {S.}~\bibnamefont
  {Deffner}}\ and\ \bibinfo {author} {\bibfnamefont {A.}~\bibnamefont
  {Saxena}},\ }\href {https://doi.org/10.1103/PhysRevLett.114.150601}
  {\bibfield  {journal} {\bibinfo  {journal} {Phys. Rev. Lett.}\ }\textbf
  {\bibinfo {volume} {114}},\ \bibinfo {pages} {150601} (\bibinfo {year}
  {2015})}\BibitemShut {NoStop}%
\bibitem [{\citenamefont {Du}\ \emph {et~al.}(2022)\citenamefont {Du},
  \citenamefont {Cao},\ and\ \citenamefont {Kou}}]{pt-finiteT}%
  \BibitemOpen
  \bibfield  {author} {\bibinfo {author} {\bibfnamefont {Q.}~\bibnamefont
  {Du}}, \bibinfo {author} {\bibfnamefont {K.}~\bibnamefont {Cao}},\ and\
  \bibinfo {author} {\bibfnamefont {S.-P.}\ \bibnamefont {Kou}},\ }\href
  {https://doi.org/10.1103/PhysRevA.106.032206} {\bibfield  {journal} {\bibinfo
   {journal} {Phys. Rev. A}\ }\textbf {\bibinfo {volume} {106}},\ \bibinfo
  {pages} {032206} (\bibinfo {year} {2022})}\BibitemShut {NoStop}%
\bibitem [{\citenamefont {Cao}\ \emph {et~al.}(2023)\citenamefont {Cao},
  \citenamefont {Du},\ and\ \citenamefont {Kou}}]{skinfiniteT}%
  \BibitemOpen
  \bibfield  {author} {\bibinfo {author} {\bibfnamefont {K.}~\bibnamefont
  {Cao}}, \bibinfo {author} {\bibfnamefont {Q.}~\bibnamefont {Du}},\ and\
  \bibinfo {author} {\bibfnamefont {S.-P.}\ \bibnamefont {Kou}},\ }\href
  {https://doi.org/10.1103/PhysRevB.108.165420} {\bibfield  {journal} {\bibinfo
   {journal} {Phys. Rev. B}\ }\textbf {\bibinfo {volume} {108}},\ \bibinfo
  {pages} {165420} (\bibinfo {year} {2023})}\BibitemShut {NoStop}%
\bibitem [{\citenamefont {Aifer}\ \emph {et~al.}(2024)\citenamefont {Aifer},
  \citenamefont {Thingna},\ and\ \citenamefont {Deffner}}]{deffner}%
  \BibitemOpen
  \bibfield  {author} {\bibinfo {author} {\bibfnamefont {M.}~\bibnamefont
  {Aifer}}, \bibinfo {author} {\bibfnamefont {J.}~\bibnamefont {Thingna}},\
  and\ \bibinfo {author} {\bibfnamefont {S.}~\bibnamefont {Deffner}},\ }\href
  {https://doi.org/10.1103/PhysRevLett.133.020401} {\bibfield  {journal}
  {\bibinfo  {journal} {Phys. Rev. Lett.}\ }\textbf {\bibinfo {volume} {133}},\
  \bibinfo {pages} {020401} (\bibinfo {year} {2024})}\BibitemShut {NoStop}%
\bibitem [{\citenamefont {Cipolloni}\ and\ \citenamefont
  {Kudler-Flam}(2024)}]{violation}%
  \BibitemOpen
  \bibfield  {author} {\bibinfo {author} {\bibfnamefont {G.}~\bibnamefont
  {Cipolloni}}\ and\ \bibinfo {author} {\bibfnamefont {J.}~\bibnamefont
  {Kudler-Flam}},\ }\href {https://doi.org/10.1103/PhysRevB.109.L020201}
  {\bibfield  {journal} {\bibinfo  {journal} {Phys. Rev. B}\ }\textbf {\bibinfo
  {volume} {109}},\ \bibinfo {pages} {L020201} (\bibinfo {year}
  {2024})}\BibitemShut {NoStop}%
\bibitem [{\citenamefont {Menczel}\ \emph {et~al.}(2024)\citenamefont
  {Menczel}, \citenamefont {Funo}, \citenamefont {Cirio}, \citenamefont
  {Lambert},\ and\ \citenamefont {Nori}}]{nori}%
  \BibitemOpen
  \bibfield  {author} {\bibinfo {author} {\bibfnamefont {P.}~\bibnamefont
  {Menczel}}, \bibinfo {author} {\bibfnamefont {K.}~\bibnamefont {Funo}},
  \bibinfo {author} {\bibfnamefont {M.}~\bibnamefont {Cirio}}, \bibinfo
  {author} {\bibfnamefont {N.}~\bibnamefont {Lambert}},\ and\ \bibinfo {author}
  {\bibfnamefont {F.}~\bibnamefont {Nori}},\ }\href
  {https://doi.org/10.1103/PhysRevResearch.6.033237} {\bibfield  {journal}
  {\bibinfo  {journal} {Phys. Rev. Res.}\ }\textbf {\bibinfo {volume} {6}},\
  \bibinfo {pages} {033237} (\bibinfo {year} {2024})}\BibitemShut {NoStop}%
\bibitem [{\citenamefont {Erdamar}\ \emph {et~al.}(2024)\citenamefont
  {Erdamar}, \citenamefont {Abbasi}, \citenamefont {Ha}, \citenamefont {Chen},
  \citenamefont {Muldoon}, \citenamefont {Joglekar},\ and\ \citenamefont
  {Murch}}]{joglekar1}%
  \BibitemOpen
  \bibfield  {author} {\bibinfo {author} {\bibfnamefont {S.}~\bibnamefont
  {Erdamar}}, \bibinfo {author} {\bibfnamefont {M.}~\bibnamefont {Abbasi}},
  \bibinfo {author} {\bibfnamefont {B.}~\bibnamefont {Ha}}, \bibinfo {author}
  {\bibfnamefont {W.}~\bibnamefont {Chen}}, \bibinfo {author} {\bibfnamefont
  {J.}~\bibnamefont {Muldoon}}, \bibinfo {author} {\bibfnamefont
  {Y.}~\bibnamefont {Joglekar}},\ and\ \bibinfo {author} {\bibfnamefont
  {K.~W.}\ \bibnamefont {Murch}},\ }\href
  {https://doi.org/10.1103/PhysRevResearch.6.L022013} {\bibfield  {journal}
  {\bibinfo  {journal} {Phys. Rev. Res.}\ }\textbf {\bibinfo {volume} {6}},\
  \bibinfo {pages} {L022013} (\bibinfo {year} {2024})}\BibitemShut {NoStop}%
\bibitem [{\citenamefont {Nishiyama}\ and\ \citenamefont
  {Hasegawa}(2025)}]{speed}%
  \BibitemOpen
  \bibfield  {author} {\bibinfo {author} {\bibfnamefont {T.}~\bibnamefont
  {Nishiyama}}\ and\ \bibinfo {author} {\bibfnamefont {Y.}~\bibnamefont
  {Hasegawa}},\ }\href {https://doi.org/10.1103/PhysRevA.111.012214} {\bibfield
   {journal} {\bibinfo  {journal} {Phys. Rev. A}\ }\textbf {\bibinfo {volume}
  {111}},\ \bibinfo {pages} {012214} (\bibinfo {year} {2025})}\BibitemShut
  {NoStop}%
\bibitem [{\citenamefont {Pino}\ \emph {et~al.}(2025)\citenamefont {Pino},
  \citenamefont {Meir},\ and\ \citenamefont {Aguado}}]{super}%
  \BibitemOpen
  \bibfield  {author} {\bibinfo {author} {\bibfnamefont {D.~M.}\ \bibnamefont
  {Pino}}, \bibinfo {author} {\bibfnamefont {Y.}~\bibnamefont {Meir}},\ and\
  \bibinfo {author} {\bibfnamefont {R.}~\bibnamefont {Aguado}},\ }\href
  {https://doi.org/10.1103/PhysRevB.111.L140503} {\bibfield  {journal}
  {\bibinfo  {journal} {Phys. Rev. B}\ }\textbf {\bibinfo {volume} {111}},\
  \bibinfo {pages} {L140503} (\bibinfo {year} {2025})}\BibitemShut {NoStop}%
\bibitem [{\citenamefont {Fring}\ and\ \citenamefont {Reboiro}(2024)}]{mraf}%
  \BibitemOpen
  \bibfield  {author} {\bibinfo {author} {\bibfnamefont {A.}~\bibnamefont
  {Fring}}\ and\ \bibinfo {author} {\bibfnamefont {M.}~\bibnamefont
  {Reboiro}},\ }\href {https://doi.org/10.1140/epjp/s13360-024-05535-y}
  {\bibfield  {journal} {\bibinfo  {journal} {The European Physical Journal
  Plus}\ }\textbf {\bibinfo {volume} {139}},\ \bibinfo {pages} {733} (\bibinfo
  {year} {2024})}\BibitemShut {NoStop}%
\bibitem [{\citenamefont {Reboiro}\ and\ \citenamefont {Tielas}(2022)}]{qrmr}%
  \BibitemOpen
  \bibfield  {author} {\bibinfo {author} {\bibfnamefont {M.}~\bibnamefont
  {Reboiro}}\ and\ \bibinfo {author} {\bibfnamefont {D.}~\bibnamefont
  {Tielas}},\ }\href {https://doi.org/10.3390/quantum4040043} {\bibfield
  {journal} {\bibinfo  {journal} {Quantum Reports}\ }\textbf {\bibinfo {volume}
  {4}},\ \bibinfo {pages} {589} (\bibinfo {year} {2022})}\BibitemShut {NoStop}%
\bibitem [{\citenamefont {Ramirez}\ and\ \citenamefont
  {Reboiro}(2022)}]{mrarxiv1}%
  \BibitemOpen
  \bibfield  {author} {\bibinfo {author} {\bibfnamefont {R.}~\bibnamefont
  {Ramirez}}\ and\ \bibinfo {author} {\bibfnamefont {M.}~\bibnamefont
  {Reboiro}},\ }\bibfield  {journal} {\bibinfo  {journal} {arXiv:2212.13173}\
  }\href {https://doi.org/10.48550/arXiv.2212.13173}
  {10.48550/arXiv.2212.13173} (\bibinfo {year} {2022})\BibitemShut {NoStop}%
\bibitem [{\citenamefont {Gao}\ \emph {et~al.}(2024)\citenamefont {Gao},
  \citenamefont {Wang}, \citenamefont {Xiao}, \citenamefont {Nakagawa},
  \citenamefont {Matsumoto}, \citenamefont {Qu}, \citenamefont {Lin},
  \citenamefont {Ueda},\ and\ \citenamefont {Xue}}]{xue}%
  \BibitemOpen
  \bibfield  {author} {\bibinfo {author} {\bibfnamefont {H.}~\bibnamefont
  {Gao}}, \bibinfo {author} {\bibfnamefont {K.}~\bibnamefont {Wang}}, \bibinfo
  {author} {\bibfnamefont {L.}~\bibnamefont {Xiao}}, \bibinfo {author}
  {\bibfnamefont {M.}~\bibnamefont {Nakagawa}}, \bibinfo {author}
  {\bibfnamefont {N.}~\bibnamefont {Matsumoto}}, \bibinfo {author}
  {\bibfnamefont {D.}~\bibnamefont {Qu}}, \bibinfo {author} {\bibfnamefont
  {H.}~\bibnamefont {Lin}}, \bibinfo {author} {\bibfnamefont {M.}~\bibnamefont
  {Ueda}},\ and\ \bibinfo {author} {\bibfnamefont {P.}~\bibnamefont {Xue}},\
  }\href {https://doi.org/10.1103/PhysRevLett.132.176601} {\bibfield  {journal}
  {\bibinfo  {journal} {Phys. Rev. Lett.}\ }\textbf {\bibinfo {volume} {132}},\
  \bibinfo {pages} {176601} (\bibinfo {year} {2024})}\BibitemShut {NoStop}%
\bibitem [{\citenamefont {Yang}\ and\ \citenamefont {Lee}(1952)}]{ylee1}%
  \BibitemOpen
  \bibfield  {author} {\bibinfo {author} {\bibfnamefont {C.~N.}\ \bibnamefont
  {Yang}}\ and\ \bibinfo {author} {\bibfnamefont {T.~D.}\ \bibnamefont {Lee}},\
  }\href {https://doi.org/10.1103/PhysRev.87.404} {\bibfield  {journal}
  {\bibinfo  {journal} {Physical Review}\ }\textbf {\bibinfo {volume} {87}},\
  \bibinfo {pages} {404} (\bibinfo {year} {1952})}\BibitemShut {NoStop}%
\bibitem [{\citenamefont {Lee}\ and\ \citenamefont {Yang}(1952)}]{ylee2}%
  \BibitemOpen
  \bibfield  {author} {\bibinfo {author} {\bibfnamefont {T.~D.}\ \bibnamefont
  {Lee}}\ and\ \bibinfo {author} {\bibfnamefont {C.~N.}\ \bibnamefont {Yang}},\
  }\href {https://doi.org/10.1103/PhysRev.87.410} {\bibfield  {journal}
  {\bibinfo  {journal} {Physical Review}\ }\textbf {\bibinfo {volume} {87}},\
  \bibinfo {pages} {410} (\bibinfo {year} {1952})}\BibitemShut {NoStop}%
\bibitem [{\citenamefont {Li}(2025)}]{ylee3}%
  \BibitemOpen
  \bibfield  {author} {\bibinfo {author} {\bibfnamefont {H.}~\bibnamefont
  {Li}},\ }\href {https://doi.org/10.1103/PhysRevB.111.045139} {\bibfield
  {journal} {\bibinfo  {journal} {Physical Review B}\ }\textbf {\bibinfo
  {volume} {111}},\ \bibinfo {pages} {045139} (\bibinfo {year}
  {2025})}\BibitemShut {NoStop}%
\bibitem [{\citenamefont {Zhu}\ \emph {et~al.}(2011)\citenamefont {Zhu} \emph
  {et~al.}}]{zhu}%
  \BibitemOpen
  \bibfield  {author} {\bibinfo {author} {\bibfnamefont {X.}~\bibnamefont
  {Zhu}} \emph {et~al.},\ }\href {https://doi.org/10.1038/nature10462}
  {\bibfield  {journal} {\bibinfo  {journal} {Nature}\ }\textbf {\bibinfo
  {volume} {478}},\ \bibinfo {pages} {211} (\bibinfo {year}
  {2011})}\BibitemShut {NoStop}%
\bibitem [{\citenamefont {L{\"u}}\ \emph {et~al.}(2013)\citenamefont {L{\"u}},
  \citenamefont {Xiang}, \citenamefont {Cui}, \citenamefont {You},\ and\
  \citenamefont {Nori}}]{zhubis}%
  \BibitemOpen
  \bibfield  {author} {\bibinfo {author} {\bibfnamefont {X.-Y.}\ \bibnamefont
  {L{\"u}}}, \bibinfo {author} {\bibfnamefont {Z.-L.}\ \bibnamefont {Xiang}},
  \bibinfo {author} {\bibfnamefont {W.}~\bibnamefont {Cui}}, \bibinfo {author}
  {\bibfnamefont {J.}~\bibnamefont {You}},\ and\ \bibinfo {author}
  {\bibfnamefont {F.}~\bibnamefont {Nori}},\ }\href
  {https://doi.org/10.1103/PhysRevA.88.012329} {\bibfield  {journal} {\bibinfo
  {journal} {Physical Review A}\ }\textbf {\bibinfo {volume} {88}},\ \bibinfo
  {pages} {012329} (\bibinfo {year} {2013})}\BibitemShut {NoStop}%
\bibitem [{\citenamefont {Marcos}\ \emph {et~al.}(2010)\citenamefont {Marcos},
  \citenamefont {Wubs}, \citenamefont {Taylor}, \citenamefont {Aguado},
  \citenamefont {Lukin},\ and\ \citenamefont {S{\o}rensen}}]{marco}%
  \BibitemOpen
  \bibfield  {author} {\bibinfo {author} {\bibfnamefont {D.}~\bibnamefont
  {Marcos}}, \bibinfo {author} {\bibfnamefont {M.}~\bibnamefont {Wubs}},
  \bibinfo {author} {\bibfnamefont {J.}~\bibnamefont {Taylor}}, \bibinfo
  {author} {\bibfnamefont {R.}~\bibnamefont {Aguado}}, \bibinfo {author}
  {\bibfnamefont {M.}~\bibnamefont {Lukin}},\ and\ \bibinfo {author}
  {\bibfnamefont {A.}~\bibnamefont {S{\o}rensen}},\ }\href
  {https://doi.org/10.1103/PhysRevLett.105.210501} {\bibfield  {journal}
  {\bibinfo  {journal} {Physical Review Letters}\ }\textbf {\bibinfo {volume}
  {105}},\ \bibinfo {pages} {210501} (\bibinfo {year} {2010})}\BibitemShut
  {NoStop}%
\bibitem [{\citenamefont {Yao}\ \emph {et~al.}(2012)\citenamefont {Yao},
  \citenamefont {Jiang}, \citenamefont {Gorshkov}, \citenamefont {Maurer},
  \citenamefont {Giedke}, \citenamefont {Cirac},\ and\ \citenamefont
  {Lukin}}]{nv-int-2}%
  \BibitemOpen
  \bibfield  {author} {\bibinfo {author} {\bibfnamefont {N.}~\bibnamefont
  {Yao}}, \bibinfo {author} {\bibfnamefont {L.}~\bibnamefont {Jiang}}, \bibinfo
  {author} {\bibfnamefont {A.}~\bibnamefont {Gorshkov}}, \bibinfo {author}
  {\bibfnamefont {P.}~\bibnamefont {Maurer}}, \bibinfo {author} {\bibfnamefont
  {G.}~\bibnamefont {Giedke}}, \bibinfo {author} {\bibfnamefont
  {J.}~\bibnamefont {Cirac}},\ and\ \bibinfo {author} {\bibfnamefont
  {M.}~\bibnamefont {Lukin}},\ }\bibfield  {journal} {\bibinfo  {journal}
  {Nature Communications}\ }\textbf {\bibinfo {volume} {3}},\ \href
  {https://doi.org/10.1038/ncomms1788} {10.1038/ncomms1788} (\bibinfo {year}
  {2012})\BibitemShut {NoStop}%
\bibitem [{\citenamefont {Doherty}\ \emph {et~al.}(2012)\citenamefont
  {Doherty}, \citenamefont {Dolde}, \citenamefont {Fedder}, \citenamefont
  {Jelezko}, \citenamefont {Wrachtrup}, \citenamefont {Manson},\ and\
  \citenamefont {Hollenberg}}]{nv-1}%
  \BibitemOpen
  \bibfield  {author} {\bibinfo {author} {\bibfnamefont {M.~W.}\ \bibnamefont
  {Doherty}}, \bibinfo {author} {\bibfnamefont {F.}~\bibnamefont {Dolde}},
  \bibinfo {author} {\bibfnamefont {H.}~\bibnamefont {Fedder}}, \bibinfo
  {author} {\bibfnamefont {F.}~\bibnamefont {Jelezko}}, \bibinfo {author}
  {\bibfnamefont {J.}~\bibnamefont {Wrachtrup}}, \bibinfo {author}
  {\bibfnamefont {N.~B.}\ \bibnamefont {Manson}},\ and\ \bibinfo {author}
  {\bibfnamefont {L.~C.~L.}\ \bibnamefont {Hollenberg}},\ }\href
  {https://doi.org/10.1103/PhysRevB.85.205203} {\bibfield  {journal} {\bibinfo
  {journal} {Phys. Rev. B}\ }\textbf {\bibinfo {volume} {85}},\ \bibinfo
  {pages} {205203} (\bibinfo {year} {2012})}\BibitemShut {NoStop}%
\bibitem [{\citenamefont {Doherty}\ \emph {et~al.}(2013)\citenamefont
  {Doherty}, \citenamefont {Manson}, \citenamefont {Delaney}, \citenamefont
  {Jelezko}, \citenamefont {Wrachtrup},\ and\ \citenamefont
  {Hollenberg}}]{nv-int-1}%
  \BibitemOpen
  \bibfield  {author} {\bibinfo {author} {\bibfnamefont {M.~W.}\ \bibnamefont
  {Doherty}}, \bibinfo {author} {\bibfnamefont {N.~B.}\ \bibnamefont {Manson}},
  \bibinfo {author} {\bibfnamefont {P.}~\bibnamefont {Delaney}}, \bibinfo
  {author} {\bibfnamefont {F.}~\bibnamefont {Jelezko}}, \bibinfo {author}
  {\bibfnamefont {J.}~\bibnamefont {Wrachtrup}},\ and\ \bibinfo {author}
  {\bibfnamefont {L.~C.}\ \bibnamefont {Hollenberg}},\ }\href
  {https://doi.org/https://doi.org/10.1016/j.physrep.2013.02.001} {\bibfield
  {journal} {\bibinfo  {journal} {Physics Reports}\ }\textbf {\bibinfo {volume}
  {528}},\ \bibinfo {pages} {1} (\bibinfo {year} {2013})},\ \bibinfo {note}
  {the nitrogen-vacancy colour centre in diamond}\BibitemShut {NoStop}%
\bibitem [{\citenamefont {Fávaro~de Oliveira}\ \emph
  {et~al.}(2017)\citenamefont {Fávaro~de Oliveira}, \citenamefont {Antonov},
  \citenamefont {Wang}, \citenamefont {Neumann}, \citenamefont {Momenzadeh},
  \citenamefont {Häußermann}, \citenamefont {Pasquarelli}, \citenamefont
  {Denisenko},\ and\ \citenamefont {Wrachtrup}}]{nv-int-new}%
  \BibitemOpen
  \bibfield  {author} {\bibinfo {author} {\bibfnamefont {F.}~\bibnamefont
  {Fávaro~de Oliveira}}, \bibinfo {author} {\bibfnamefont {D.}~\bibnamefont
  {Antonov}}, \bibinfo {author} {\bibfnamefont {Y.}~\bibnamefont {Wang}},
  \bibinfo {author} {\bibfnamefont {P.}~\bibnamefont {Neumann}}, \bibinfo
  {author} {\bibfnamefont {S.~A.}\ \bibnamefont {Momenzadeh}}, \bibinfo
  {author} {\bibfnamefont {T.}~\bibnamefont {Häußermann}}, \bibinfo {author}
  {\bibfnamefont {A.}~\bibnamefont {Pasquarelli}}, \bibinfo {author}
  {\bibfnamefont {A.}~\bibnamefont {Denisenko}},\ and\ \bibinfo {author}
  {\bibfnamefont {J.}~\bibnamefont {Wrachtrup}},\ }\bibfield  {journal}
  {\bibinfo  {journal} {Nature Communications}\ }\textbf {\bibinfo {volume}
  {8}},\ \href {https://doi.org/10.1038/ncomms15409} {10.1038/ncomms15409}
  (\bibinfo {year} {2017})\BibitemShut {NoStop}%
\bibitem [{\citenamefont {Ali}\ \emph {et~al.}(2018)\citenamefont {Ali},
  \citenamefont {Basit}, \citenamefont {Badshah},\ and\ \citenamefont
  {Ge}}]{hybrid-10}%
  \BibitemOpen
  \bibfield  {author} {\bibinfo {author} {\bibfnamefont {H.}~\bibnamefont
  {Ali}}, \bibinfo {author} {\bibfnamefont {A.}~\bibnamefont {Basit}}, \bibinfo
  {author} {\bibfnamefont {F.}~\bibnamefont {Badshah}},\ and\ \bibinfo {author}
  {\bibfnamefont {G.-Q.}\ \bibnamefont {Ge}},\ }\href
  {https://doi.org/10.1016/j.physe.2018.07.040} {\bibfield  {journal} {\bibinfo
   {journal} {Physica E}\ }\textbf {\bibinfo {volume} {104}},\ \bibinfo {pages}
  {261} (\bibinfo {year} {2018})}\BibitemShut {NoStop}%
\bibitem [{\citenamefont {Albrecht}\ \emph {et~al.}(2014)\citenamefont
  {Albrecht} \emph {et~al.}}]{hybrid-11}%
  \BibitemOpen
  \bibfield  {author} {\bibinfo {author} {\bibfnamefont {A.}~\bibnamefont
  {Albrecht}} \emph {et~al.},\ }\href
  {https://doi.org/10.1088/1367-2630/16/9/093002} {\bibfield  {journal}
  {\bibinfo  {journal} {New Journal of Physics}\ }\textbf {\bibinfo {volume}
  {16}},\ \bibinfo {pages} {093002} (\bibinfo {year} {2014})}\BibitemShut
  {NoStop}%
\bibitem [{\citenamefont {Reboiro}\ \emph {et~al.}(2017)\citenamefont
  {Reboiro}, \citenamefont {Civitarese},\ and\ \citenamefont
  {Ram{\'\i}rez}}]{nosap17}%
  \BibitemOpen
  \bibfield  {author} {\bibinfo {author} {\bibfnamefont {M.}~\bibnamefont
  {Reboiro}}, \bibinfo {author} {\bibfnamefont {O.}~\bibnamefont
  {Civitarese}},\ and\ \bibinfo {author} {\bibfnamefont {R.}~\bibnamefont
  {Ram{\'\i}rez}},\ }\href {https://doi.org/10.1016/j.aop.2017.01.025}
  {\bibfield  {journal} {\bibinfo  {journal} {Annals of Physics}\ }\textbf
  {\bibinfo {volume} {378}},\ \bibinfo {pages} {418} (\bibinfo {year}
  {2017})}\BibitemShut {NoStop}%
\bibitem [{\citenamefont {Ram{\'\i}rez}\ \emph {et~al.}(2020)\citenamefont
  {Ram{\'\i}rez}, \citenamefont {Reboiro},\ and\ \citenamefont
  {Tielas}}]{epjd}%
  \BibitemOpen
  \bibfield  {author} {\bibinfo {author} {\bibfnamefont {R.}~\bibnamefont
  {Ram{\'\i}rez}}, \bibinfo {author} {\bibfnamefont {M.}~\bibnamefont
  {Reboiro}},\ and\ \bibinfo {author} {\bibfnamefont {D.}~\bibnamefont
  {Tielas}},\ }\href@noop {} {\bibfield  {journal} {\bibinfo  {journal} {The
  European Physical Journal D}\ }\textbf {\bibinfo {volume} {74}},\ \bibinfo
  {pages} {193} (\bibinfo {year} {2020})}\BibitemShut {NoStop}%
\bibitem [{\citenamefont {Cambria}\ \emph {et~al.}(2023)\citenamefont
  {Cambria}, \citenamefont {Thiering}, \citenamefont {Norambuena},
  \citenamefont {Dinani}, \citenamefont {Gardill}, \citenamefont {Kemeny},
  \citenamefont {Lordi}, \citenamefont {Gali}, \citenamefont {Maze},\ and\
  \citenamefont {Kolkowitz}}]{nvsT}%
  \BibitemOpen
  \bibfield  {author} {\bibinfo {author} {\bibfnamefont {M.~C.}\ \bibnamefont
  {Cambria}}, \bibinfo {author} {\bibfnamefont {G.}~\bibnamefont {Thiering}},
  \bibinfo {author} {\bibfnamefont {A.}~\bibnamefont {Norambuena}}, \bibinfo
  {author} {\bibfnamefont {H.~T.}\ \bibnamefont {Dinani}}, \bibinfo {author}
  {\bibfnamefont {A.}~\bibnamefont {Gardill}}, \bibinfo {author} {\bibfnamefont
  {I.}~\bibnamefont {Kemeny}}, \bibinfo {author} {\bibfnamefont
  {V.}~\bibnamefont {Lordi}}, \bibinfo {author} {\bibfnamefont
  {A.}~\bibnamefont {Gali}}, \bibinfo {author} {\bibfnamefont {J.~R.}\
  \bibnamefont {Maze}},\ and\ \bibinfo {author} {\bibfnamefont
  {S.}~\bibnamefont {Kolkowitz}},\ }\href
  {https://doi.org/10.1103/PhysRevB.108.L180102} {\bibfield  {journal}
  {\bibinfo  {journal} {Phys. Rev. B}\ }\textbf {\bibinfo {volume} {108}},\
  \bibinfo {pages} {L180102} (\bibinfo {year} {2023})}\BibitemShut {NoStop}%
\bibitem [{\citenamefont {Ring}\ and\ \citenamefont {Schuck}(1980)}]{ring}%
  \BibitemOpen
  \bibfield  {author} {\bibinfo {author} {\bibfnamefont {P.}~\bibnamefont
  {Ring}}\ and\ \bibinfo {author} {\bibfnamefont {P.}~\bibnamefont {Schuck}},\
  }\bibinfo {title} {The nuclear many-body problem}\ (\bibinfo  {publisher}
  {Springer{-}Verlag},\ \bibinfo {address} {New York},\ \bibinfo {year}
  {1980})\BibitemShut {NoStop}%
\bibitem [{\citenamefont {Martinis}(2004)}]{sfqbT3}%
  \BibitemOpen
  \bibfield  {author} {\bibinfo {author} {\bibfnamefont {J.~M.}\ \bibnamefont
  {Martinis}}\ }(\bibinfo  {publisher} {Elsevier},\ \bibinfo {year} {2004})\
  pp.\ \bibinfo {pages} {487--520}\BibitemShut {NoStop}%
\bibitem [{\citenamefont {Anderson}(1958)}]{anderson}%
  \BibitemOpen
  \bibfield  {author} {\bibinfo {author} {\bibfnamefont {P.~W.}\ \bibnamefont
  {Anderson}},\ }\href {https://doi.org/10.1103/PhysRev.112.1900} {\bibfield
  {journal} {\bibinfo  {journal} {Phys. Rev.}\ }\textbf {\bibinfo {volume}
  {112}},\ \bibinfo {pages} {1900} (\bibinfo {year} {1958})}\BibitemShut
  {NoStop}%
\bibitem [{\citenamefont {Bardeen}\ \emph {et~al.}(1957)\citenamefont
  {Bardeen}, \citenamefont {Cooper},\ and\ \citenamefont {Schrieffer}}]{bcs}%
  \BibitemOpen
  \bibfield  {author} {\bibinfo {author} {\bibfnamefont {J.}~\bibnamefont
  {Bardeen}}, \bibinfo {author} {\bibfnamefont {L.~N.}\ \bibnamefont
  {Cooper}},\ and\ \bibinfo {author} {\bibfnamefont {J.~R.}\ \bibnamefont
  {Schrieffer}},\ }\href {https://doi.org/10.1103/PhysRev.108.1175} {\bibfield
  {journal} {\bibinfo  {journal} {Phys. Rev.}\ }\textbf {\bibinfo {volume}
  {108}},\ \bibinfo {pages} {1175} (\bibinfo {year} {1957})}\BibitemShut
  {NoStop}%
\bibitem [{\citenamefont {Feynman}(1972)}]{feynman}%
  \BibitemOpen
  \bibfield  {author} {\bibinfo {author} {\bibfnamefont {R.~P.}\ \bibnamefont
  {Feynman}},\ }\bibinfo {title} {Statistical mechanics. a set of lectures}\
  (\bibinfo  {publisher} {W. A. Benjamin Inc.},\ \bibinfo {year}
  {1972})\BibitemShut {NoStop}%
\bibitem [{\citenamefont {Saitoh}\ \emph {et~al.}(1998)\citenamefont {Saitoh},
  \citenamefont {Utagawa},\ and\ \citenamefont {Enomoto}}]{sfqbT1}%
  \BibitemOpen
  \bibfield  {author} {\bibinfo {author} {\bibfnamefont {K.}~\bibnamefont
  {Saitoh}}, \bibinfo {author} {\bibfnamefont {T.}~\bibnamefont {Utagawa}},\
  and\ \bibinfo {author} {\bibfnamefont {Y.}~\bibnamefont {Enomoto}},\ }\href
  {https://doi.org/10.1063/1.121080} {\bibfield  {journal} {\bibinfo  {journal}
  {Applied Physics Letters}\ }\textbf {\bibinfo {volume} {72}},\ \bibinfo
  {pages} {2754} (\bibinfo {year} {1998})}\BibitemShut {NoStop}%
\bibitem [{\citenamefont {Khare}\ \emph {et~al.}(1999)\citenamefont {Khare},
  \citenamefont {Gupta}, \citenamefont {Hossain}, \citenamefont {Nagarajan},
  \citenamefont {Gupta},\ and\ \citenamefont {Vijayaraghavan}}]{sfqbT2}%
  \BibitemOpen
  \bibfield  {author} {\bibinfo {author} {\bibfnamefont {N.}~\bibnamefont
  {Khare}}, \bibinfo {author} {\bibfnamefont {A.}~\bibnamefont {Gupta}},
  \bibinfo {author} {\bibfnamefont {Z.}~\bibnamefont {Hossain}}, \bibinfo
  {author} {\bibfnamefont {R.}~\bibnamefont {Nagarajan}}, \bibinfo {author}
  {\bibfnamefont {L.}~\bibnamefont {Gupta}},\ and\ \bibinfo {author}
  {\bibfnamefont {R.}~\bibnamefont {Vijayaraghavan}},\ }\href
  {https://doi.org/https://doi.org/10.1016/S0921-4534(99)00195-1} {\bibfield
  {journal} {\bibinfo  {journal} {Physica C: Superconductivity}\ }\textbf
  {\bibinfo {volume} {316}},\ \bibinfo {pages} {257} (\bibinfo {year}
  {1999})}\BibitemShut {NoStop}%
\bibitem [{\citenamefont {Valenzuela}\ \emph {et~al.}(2006)\citenamefont
  {Valenzuela}, \citenamefont {Oliver}, \citenamefont {Berns}, \citenamefont
  {Berggren}, \citenamefont {Levitov},\ and\ \citenamefont {Orlando}}]{sfqbT4}%
  \BibitemOpen
  \bibfield  {author} {\bibinfo {author} {\bibfnamefont {S.~O.}\ \bibnamefont
  {Valenzuela}}, \bibinfo {author} {\bibfnamefont {W.~D.}\ \bibnamefont
  {Oliver}}, \bibinfo {author} {\bibfnamefont {D.~M.}\ \bibnamefont {Berns}},
  \bibinfo {author} {\bibfnamefont {K.~K.}\ \bibnamefont {Berggren}}, \bibinfo
  {author} {\bibfnamefont {L.~S.}\ \bibnamefont {Levitov}},\ and\ \bibinfo
  {author} {\bibfnamefont {T.~P.}\ \bibnamefont {Orlando}},\ }\href
  {https://doi.org/10.1126/science.1134008} {\bibfield  {journal} {\bibinfo
  {journal} {Science}\ }\textbf {\bibinfo {volume} {314}},\ \bibinfo {pages}
  {1589} (\bibinfo {year} {2006})},\ \Eprint
  {https://arxiv.org/abs/https://www.science.org/doi/pdf/10.1126/science.1134008}
  {https://www.science.org/doi/pdf/10.1126/science.1134008} \BibitemShut
  {NoStop}%
\bibitem [{\citenamefont {Lisenfeld}\ \emph {et~al.}(2007)\citenamefont
  {Lisenfeld}, \citenamefont {Lukashenko}, \citenamefont {Ansmann},
  \citenamefont {Martinis},\ and\ \citenamefont {Ustinov}}]{sfqbT5}%
  \BibitemOpen
  \bibfield  {author} {\bibinfo {author} {\bibfnamefont {J.}~\bibnamefont
  {Lisenfeld}}, \bibinfo {author} {\bibfnamefont {A.}~\bibnamefont
  {Lukashenko}}, \bibinfo {author} {\bibfnamefont {M.}~\bibnamefont {Ansmann}},
  \bibinfo {author} {\bibfnamefont {J.~M.}\ \bibnamefont {Martinis}},\ and\
  \bibinfo {author} {\bibfnamefont {A.~V.}\ \bibnamefont {Ustinov}},\ }\href
  {https://doi.org/10.1103/PhysRevLett.99.170504} {\bibfield  {journal}
  {\bibinfo  {journal} {Phys. Rev. Lett.}\ }\textbf {\bibinfo {volume} {99}},\
  \bibinfo {pages} {170504} (\bibinfo {year} {2007})}\BibitemShut {NoStop}%
\bibitem [{\citenamefont {Wenner}\ \emph {et~al.}(2013)\citenamefont {Wenner},
  \citenamefont {Yin}, \citenamefont {Lucero}, \citenamefont {Barends},
  \citenamefont {Chen}, \citenamefont {Chiaro}, \citenamefont {Kelly},
  \citenamefont {Lenander}, \citenamefont {Mariantoni}, \citenamefont
  {Megrant}, \citenamefont {Neill}, \citenamefont {O'Malley}, \citenamefont
  {Sank}, \citenamefont {Vainsencher}, \citenamefont {Wang}, \citenamefont
  {White}, \citenamefont {Cleland},\ and\ \citenamefont {Martinis}}]{sfqbT6}%
  \BibitemOpen
  \bibfield  {author} {\bibinfo {author} {\bibfnamefont {J.}~\bibnamefont
  {Wenner}}, \bibinfo {author} {\bibfnamefont {Y.}~\bibnamefont {Yin}},
  \bibinfo {author} {\bibfnamefont {E.}~\bibnamefont {Lucero}}, \bibinfo
  {author} {\bibfnamefont {R.}~\bibnamefont {Barends}}, \bibinfo {author}
  {\bibfnamefont {Y.}~\bibnamefont {Chen}}, \bibinfo {author} {\bibfnamefont
  {B.}~\bibnamefont {Chiaro}}, \bibinfo {author} {\bibfnamefont
  {J.}~\bibnamefont {Kelly}}, \bibinfo {author} {\bibfnamefont
  {M.}~\bibnamefont {Lenander}}, \bibinfo {author} {\bibfnamefont
  {M.}~\bibnamefont {Mariantoni}}, \bibinfo {author} {\bibfnamefont
  {A.}~\bibnamefont {Megrant}}, \bibinfo {author} {\bibfnamefont
  {C.}~\bibnamefont {Neill}}, \bibinfo {author} {\bibfnamefont {P.~J.~J.}\
  \bibnamefont {O'Malley}}, \bibinfo {author} {\bibfnamefont {D.}~\bibnamefont
  {Sank}}, \bibinfo {author} {\bibfnamefont {A.}~\bibnamefont {Vainsencher}},
  \bibinfo {author} {\bibfnamefont {H.}~\bibnamefont {Wang}}, \bibinfo {author}
  {\bibfnamefont {T.~C.}\ \bibnamefont {White}}, \bibinfo {author}
  {\bibfnamefont {A.~N.}\ \bibnamefont {Cleland}},\ and\ \bibinfo {author}
  {\bibfnamefont {J.~M.}\ \bibnamefont {Martinis}},\ }\href
  {https://doi.org/10.1103/PhysRevLett.110.150502} {\bibfield  {journal}
  {\bibinfo  {journal} {Phys. Rev. Lett.}\ }\textbf {\bibinfo {volume} {110}},\
  \bibinfo {pages} {150502} (\bibinfo {year} {2013})}\BibitemShut {NoStop}%
\bibitem [{\citenamefont {Spilla}\ \emph {et~al.}(2014)\citenamefont {Spilla},
  \citenamefont {Hassler},\ and\ \citenamefont {Splettstoesser}}]{sfqbT7}%
  \BibitemOpen
  \bibfield  {author} {\bibinfo {author} {\bibfnamefont {S.}~\bibnamefont
  {Spilla}}, \bibinfo {author} {\bibfnamefont {F.}~\bibnamefont {Hassler}},\
  and\ \bibinfo {author} {\bibfnamefont {J.}~\bibnamefont {Splettstoesser}},\
  }\href {https://doi.org/10.1088/1367-2630/16/4/045020} {\bibfield  {journal}
  {\bibinfo  {journal} {New Journal of Physics}\ }\textbf {\bibinfo {volume}
  {16}},\ \bibinfo {pages} {045020} (\bibinfo {year} {2014})}\BibitemShut
  {NoStop}%
\bibitem [{\citenamefont {Kakuyanagi}\ \emph {et~al.}(2023)\citenamefont
  {Kakuyanagi}, \citenamefont {Toida}, \citenamefont {Abdurakhimov},\ and\
  \citenamefont {Saito}}]{sfqbT8}%
  \BibitemOpen
  \bibfield  {author} {\bibinfo {author} {\bibfnamefont {K.}~\bibnamefont
  {Kakuyanagi}}, \bibinfo {author} {\bibfnamefont {H.}~\bibnamefont {Toida}},
  \bibinfo {author} {\bibfnamefont {L.~V.}\ \bibnamefont {Abdurakhimov}},\ and\
  \bibinfo {author} {\bibfnamefont {S.}~\bibnamefont {Saito}},\ }\href
  {https://doi.org/10.1088/1367-2630/acb379} {\bibfield  {journal} {\bibinfo
  {journal} {New Journal of Physics}\ }\textbf {\bibinfo {volume} {25}},\
  \bibinfo {pages} {013036} (\bibinfo {year} {2023})}\BibitemShut {NoStop}%
\bibitem [{\citenamefont {Mizuochi}\ \emph {et~al.}(2009)\citenamefont
  {Mizuochi}, \citenamefont {Neumann}, \citenamefont {Rempp}, \citenamefont
  {Beck}, \citenamefont {Jacques}, \citenamefont {Siyushev}, \citenamefont
  {Nakamura}, \citenamefont {Twitchen}, \citenamefont {Watanabe}, \citenamefont
  {Yamasaki}, \citenamefont {Jelezko},\ and\ \citenamefont
  {Wrachtrup}}]{nv-ct1}%
  \BibitemOpen
  \bibfield  {author} {\bibinfo {author} {\bibfnamefont {N.}~\bibnamefont
  {Mizuochi}}, \bibinfo {author} {\bibfnamefont {P.}~\bibnamefont {Neumann}},
  \bibinfo {author} {\bibfnamefont {F.}~\bibnamefont {Rempp}}, \bibinfo
  {author} {\bibfnamefont {J.}~\bibnamefont {Beck}}, \bibinfo {author}
  {\bibfnamefont {V.}~\bibnamefont {Jacques}}, \bibinfo {author} {\bibfnamefont
  {P.}~\bibnamefont {Siyushev}}, \bibinfo {author} {\bibfnamefont
  {K.}~\bibnamefont {Nakamura}}, \bibinfo {author} {\bibfnamefont {D.~J.}\
  \bibnamefont {Twitchen}}, \bibinfo {author} {\bibfnamefont {H.}~\bibnamefont
  {Watanabe}}, \bibinfo {author} {\bibfnamefont {S.}~\bibnamefont {Yamasaki}},
  \bibinfo {author} {\bibfnamefont {F.}~\bibnamefont {Jelezko}},\ and\ \bibinfo
  {author} {\bibfnamefont {J.}~\bibnamefont {Wrachtrup}},\ }\href
  {https://doi.org/10.1103/PhysRevB.80.041201} {\bibfield  {journal} {\bibinfo
  {journal} {Phys. Rev. B}\ }\textbf {\bibinfo {volume} {80}},\ \bibinfo
  {pages} {041201} (\bibinfo {year} {2009})}\BibitemShut {NoStop}%
\bibitem [{\citenamefont {Stanwix}\ \emph {et~al.}(2010)\citenamefont
  {Stanwix}, \citenamefont {Pham}, \citenamefont {Maze}, \citenamefont
  {Le~Sage}, \citenamefont {Yeung}, \citenamefont {Cappellaro}, \citenamefont
  {Hemmer}, \citenamefont {Yacoby}, \citenamefont {Lukin},\ and\ \citenamefont
  {Walsworth}}]{nv-ct2}%
  \BibitemOpen
  \bibfield  {author} {\bibinfo {author} {\bibfnamefont {P.~L.}\ \bibnamefont
  {Stanwix}}, \bibinfo {author} {\bibfnamefont {L.~M.}\ \bibnamefont {Pham}},
  \bibinfo {author} {\bibfnamefont {J.~R.}\ \bibnamefont {Maze}}, \bibinfo
  {author} {\bibfnamefont {D.}~\bibnamefont {Le~Sage}}, \bibinfo {author}
  {\bibfnamefont {T.~K.}\ \bibnamefont {Yeung}}, \bibinfo {author}
  {\bibfnamefont {P.}~\bibnamefont {Cappellaro}}, \bibinfo {author}
  {\bibfnamefont {P.~R.}\ \bibnamefont {Hemmer}}, \bibinfo {author}
  {\bibfnamefont {A.}~\bibnamefont {Yacoby}}, \bibinfo {author} {\bibfnamefont
  {M.~D.}\ \bibnamefont {Lukin}},\ and\ \bibinfo {author} {\bibfnamefont
  {R.~L.}\ \bibnamefont {Walsworth}},\ }\href
  {https://doi.org/10.1103/PhysRevB.82.201201} {\bibfield  {journal} {\bibinfo
  {journal} {Phys. Rev. B}\ }\textbf {\bibinfo {volume} {82}},\ \bibinfo
  {pages} {201201} (\bibinfo {year} {2010})}\BibitemShut {NoStop}%
\bibitem [{\citenamefont {Gulka}\ \emph {et~al.}(2017)\citenamefont {Gulka},
  \citenamefont {Bourgeois}, \citenamefont {Hruby}, \citenamefont {Siyushev},
  \citenamefont {Wachter}, \citenamefont {Aumayr}, \citenamefont {Hemmer},
  \citenamefont {Gali}, \citenamefont {Jelezko}, \citenamefont {Trupke},\ and\
  \citenamefont {Nesladek}}]{nv-ct3}%
  \BibitemOpen
  \bibfield  {author} {\bibinfo {author} {\bibfnamefont {M.}~\bibnamefont
  {Gulka}}, \bibinfo {author} {\bibfnamefont {E.}~\bibnamefont {Bourgeois}},
  \bibinfo {author} {\bibfnamefont {J.}~\bibnamefont {Hruby}}, \bibinfo
  {author} {\bibfnamefont {P.}~\bibnamefont {Siyushev}}, \bibinfo {author}
  {\bibfnamefont {G.}~\bibnamefont {Wachter}}, \bibinfo {author} {\bibfnamefont
  {F.}~\bibnamefont {Aumayr}}, \bibinfo {author} {\bibfnamefont {P.~R.}\
  \bibnamefont {Hemmer}}, \bibinfo {author} {\bibfnamefont {A.}~\bibnamefont
  {Gali}}, \bibinfo {author} {\bibfnamefont {F.}~\bibnamefont {Jelezko}},
  \bibinfo {author} {\bibfnamefont {M.}~\bibnamefont {Trupke}},\ and\ \bibinfo
  {author} {\bibfnamefont {M.}~\bibnamefont {Nesladek}},\ }\href
  {https://doi.org/10.1103/PhysRevApplied.7.044032} {\bibfield  {journal}
  {\bibinfo  {journal} {Phys. Rev. Appl.}\ }\textbf {\bibinfo {volume} {7}},\
  \bibinfo {pages} {044032} (\bibinfo {year} {2017})}\BibitemShut {NoStop}%
\bibitem [{\citenamefont {van Wyk}\ \emph {et~al.}(1997)\citenamefont {van
  Wyk}, \citenamefont {Reynhardt}, \citenamefont {High},\ and\ \citenamefont
  {Kiflawi}}]{linew1}%
  \BibitemOpen
  \bibfield  {author} {\bibinfo {author} {\bibfnamefont {J.~A.}\ \bibnamefont
  {van Wyk}}, \bibinfo {author} {\bibfnamefont {E.~C.}\ \bibnamefont
  {Reynhardt}}, \bibinfo {author} {\bibfnamefont {G.~L.}\ \bibnamefont
  {High}},\ and\ \bibinfo {author} {\bibfnamefont {I.}~\bibnamefont
  {Kiflawi}},\ }\href {https://doi.org/10.1088/0022-3727/30/12/016} {\bibfield
  {journal} {\bibinfo  {journal} {Journal of Physics D: Applied Physics}\
  }\textbf {\bibinfo {volume} {30}},\ \bibinfo {pages} {1790} (\bibinfo {year}
  {1997})}\BibitemShut {NoStop}%
\bibitem [{\citenamefont {Stepanov}\ and\ \citenamefont
  {Takahashi}(2016)}]{linew2}%
  \BibitemOpen
  \bibfield  {author} {\bibinfo {author} {\bibfnamefont {V.}~\bibnamefont
  {Stepanov}}\ and\ \bibinfo {author} {\bibfnamefont {S.}~\bibnamefont
  {Takahashi}},\ }\href {https://doi.org/10.1103/PhysRevB.94.024421} {\bibfield
   {journal} {\bibinfo  {journal} {Phys. Rev. B}\ }\textbf {\bibinfo {volume}
  {94}},\ \bibinfo {pages} {024421} (\bibinfo {year} {2016})}\BibitemShut
  {NoStop}%
\bibitem [{\citenamefont {Clements}\ \emph {et~al.}(2018)\citenamefont
  {Clements}, \citenamefont {Renema}, \citenamefont {Eckstein}, \citenamefont
  {Valido}, \citenamefont {Lita}, \citenamefont {Gerrits}, \citenamefont {Nam},
  \citenamefont {Kolthammer}, \citenamefont {Huh},\ and\ \citenamefont
  {Walmsley}}]{asymmetry}%
  \BibitemOpen
  \bibfield  {author} {\bibinfo {author} {\bibfnamefont {W.~R.}\ \bibnamefont
  {Clements}}, \bibinfo {author} {\bibfnamefont {J.~J.}\ \bibnamefont
  {Renema}}, \bibinfo {author} {\bibfnamefont {A.}~\bibnamefont {Eckstein}},
  \bibinfo {author} {\bibfnamefont {A.~A.}\ \bibnamefont {Valido}}, \bibinfo
  {author} {\bibfnamefont {A.}~\bibnamefont {Lita}}, \bibinfo {author}
  {\bibfnamefont {T.}~\bibnamefont {Gerrits}}, \bibinfo {author} {\bibfnamefont
  {S.~W.}\ \bibnamefont {Nam}}, \bibinfo {author} {\bibfnamefont {W.~S.}\
  \bibnamefont {Kolthammer}}, \bibinfo {author} {\bibfnamefont
  {J.}~\bibnamefont {Huh}},\ and\ \bibinfo {author} {\bibfnamefont {I.~A.}\
  \bibnamefont {Walmsley}},\ }\href {https://doi.org/10.1088/1361-6455/aaf031}
  {\bibfield  {journal} {\bibinfo  {journal} {Journal of Physics B: Atomic,
  Molecular and Optical Physics}\ }\textbf {\bibinfo {volume} {51}},\ \bibinfo
  {pages} {245503} (\bibinfo {year} {2018})}\BibitemShut {NoStop}%
\bibitem [{\citenamefont {Wu}\ \emph {et~al.}(2019)\citenamefont {Wu},
  \citenamefont {Liu}, \citenamefont {Geng}, \citenamefont {Song},
  \citenamefont {Ye}, \citenamefont {Duan}, \citenamefont {Rong},\ and\
  \citenamefont {Du}}]{asymmetry1}%
  \BibitemOpen
  \bibfield  {author} {\bibinfo {author} {\bibfnamefont {Y.}~\bibnamefont
  {Wu}}, \bibinfo {author} {\bibfnamefont {W.}~\bibnamefont {Liu}}, \bibinfo
  {author} {\bibfnamefont {J.}~\bibnamefont {Geng}}, \bibinfo {author}
  {\bibfnamefont {X.}~\bibnamefont {Song}}, \bibinfo {author} {\bibfnamefont
  {X.}~\bibnamefont {Ye}}, \bibinfo {author} {\bibfnamefont {C.-K.}\
  \bibnamefont {Duan}}, \bibinfo {author} {\bibfnamefont {X.}~\bibnamefont
  {Rong}},\ and\ \bibinfo {author} {\bibfnamefont {J.}~\bibnamefont {Du}},\
  }\href {https://doi.org/10.1126/science.aaw8205} {\bibfield  {journal}
  {\bibinfo  {journal} {Science}\ }\textbf {\bibinfo {volume} {364}},\ \bibinfo
  {pages} {878} (\bibinfo {year} {2019})},\ \Eprint
  {https://arxiv.org/abs/https://www.science.org/doi/pdf/10.1126/science.aaw8205}
  {https://www.science.org/doi/pdf/10.1126/science.aaw8205} \BibitemShut
  {NoStop}%
\bibitem [{\citenamefont {Ma}\ \emph {et~al.}(2013)\citenamefont {Ma},
  \citenamefont {Li}, \citenamefont {Fang}, \citenamefont {Gao},\ and\
  \citenamefont {Li}}]{asymmetry2}%
  \BibitemOpen
  \bibfield  {author} {\bibinfo {author} {\bibfnamefont {S.-l.}\ \bibnamefont
  {Ma}}, \bibinfo {author} {\bibfnamefont {P.-b.}\ \bibnamefont {Li}}, \bibinfo
  {author} {\bibfnamefont {A.-p.}\ \bibnamefont {Fang}}, \bibinfo {author}
  {\bibfnamefont {S.-y.}\ \bibnamefont {Gao}},\ and\ \bibinfo {author}
  {\bibfnamefont {F.-l.}\ \bibnamefont {Li}},\ }\href
  {https://doi.org/10.1103/PhysRevA.88.013837} {\bibfield  {journal} {\bibinfo
  {journal} {Phys. Rev. A}\ }\textbf {\bibinfo {volume} {88}},\ \bibinfo
  {pages} {013837} (\bibinfo {year} {2013})}\BibitemShut {NoStop}%
\bibitem [{\citenamefont {Ram{\'\i}rez}\ and\ \citenamefont
  {Reboiro}(2019)}]{jmp19}%
  \BibitemOpen
  \bibfield  {author} {\bibinfo {author} {\bibfnamefont {R.}~\bibnamefont
  {Ram{\'\i}rez}}\ and\ \bibinfo {author} {\bibfnamefont {M.}~\bibnamefont
  {Reboiro}},\ }\href {https://doi.org/10.1063/1.5075628} {\bibfield  {journal}
  {\bibinfo  {journal} {Journal of Mathematical Physics}\ }\textbf {\bibinfo
  {volume} {60}},\ \bibinfo {pages} {012106} (\bibinfo {year}
  {2019})}\BibitemShut {NoStop}%
\bibitem [{\citenamefont {Vdovin}\ and\ \citenamefont
  {Storozhenko}(1999)}]{dapro-lipkin}%
  \BibitemOpen
  \bibfield  {author} {\bibinfo {author} {\bibfnamefont {A.}~\bibnamefont
  {Vdovin}}\ and\ \bibinfo {author} {\bibfnamefont {A.}~\bibnamefont
  {Storozhenko}},\ }\href@noop {} {\bibfield  {journal} {\bibinfo  {journal}
  {The European Physical Journal A-Hadrons and Nuclei}\ }\textbf {\bibinfo
  {volume} {5}},\ \bibinfo {pages} {263} (\bibinfo {year} {1999})}\BibitemShut
  {NoStop}%
\bibitem [{\citenamefont {Kuriyama}\ \emph {et~al.}(2002)\citenamefont
  {Kuriyama}, \citenamefont {Providência}, \citenamefont {Tsue},\ and\
  \citenamefont {Yamamura}}]{dapro-pair}%
  \BibitemOpen
  \bibfield  {author} {\bibinfo {author} {\bibfnamefont {A.}~\bibnamefont
  {Kuriyama}}, \bibinfo {author} {\bibfnamefont {J.~d.}\ \bibnamefont
  {Providência}}, \bibinfo {author} {\bibfnamefont {Y.}~\bibnamefont {Tsue}},\
  and\ \bibinfo {author} {\bibfnamefont {M.}~\bibnamefont {Yamamura}},\ }\href
  {https://doi.org/10.1143/PTP.107.43} {\bibfield  {journal} {\bibinfo
  {journal} {Progress of Theoretical Physics}\ }\textbf {\bibinfo {volume}
  {107}},\ \bibinfo {pages} {43} (\bibinfo {year} {2002})},\ \Eprint
  {https://arxiv.org/abs/https://academic.oup.com/ptp/article-pdf/107/1/43/5213198/107-1-43.pdf}
  {https://academic.oup.com/ptp/article-pdf/107/1/43/5213198/107-1-43.pdf}
  \BibitemShut {NoStop}%
\bibitem [{\citenamefont {Cahn}\ and\ \citenamefont
  {Hilliard}(1958)}]{spinodal}%
  \BibitemOpen
  \bibfield  {author} {\bibinfo {author} {\bibfnamefont {J.~W.}\ \bibnamefont
  {Cahn}}\ and\ \bibinfo {author} {\bibfnamefont {J.~E.}\ \bibnamefont
  {Hilliard}},\ }\href {https://doi.org/10.1063/1.1744102} {\bibfield
  {journal} {\bibinfo  {journal} {The Journal of Chemical Physics}\ }\textbf
  {\bibinfo {volume} {28}},\ \bibinfo {pages} {258} (\bibinfo {year}
  {1958})}\BibitemShut {NoStop}%
\bibitem [{\citenamefont {Klein}\ and\ \citenamefont
  {Marshalek}(1991)}]{klein1991boson}%
  \BibitemOpen
  \bibfield  {author} {\bibinfo {author} {\bibfnamefont {A.}~\bibnamefont
  {Klein}}\ and\ \bibinfo {author} {\bibfnamefont {E.}~\bibnamefont
  {Marshalek}},\ }\href@noop {} {\bibfield  {journal} {\bibinfo  {journal}
  {Reviews of modern physics}\ }\textbf {\bibinfo {volume} {63}},\ \bibinfo
  {pages} {375} (\bibinfo {year} {1991})}\BibitemShut {NoStop}%
\bibitem [{\citenamefont {Jarzynski}\ and\ \citenamefont
  {W\'ojcik}(2004)}]{jarzynski2}%
  \BibitemOpen
  \bibfield  {author} {\bibinfo {author} {\bibfnamefont {C.}~\bibnamefont
  {Jarzynski}}\ and\ \bibinfo {author} {\bibfnamefont {D.~K.}\ \bibnamefont
  {W\'ojcik}},\ }\href {https://doi.org/10.1103/PhysRevLett.92.230602}
  {\bibfield  {journal} {\bibinfo  {journal} {Phys. Rev. Lett.}\ }\textbf
  {\bibinfo {volume} {92}},\ \bibinfo {pages} {230602} (\bibinfo {year}
  {2004})}\BibitemShut {NoStop}%
\end{thebibliography}
\end{document}